%% file: main.tex
\documentclass[10pt,twocolumn,letterpaper]{article}

\usepackage[pagenumbers]{cvpr} % To force page numbers, e.g. for an arXiv version
\usepackage[accsupp]{axessibility} % Improves PDF readability for those with visual impairments.
\usepackage{adjustbox}
\usepackage{algorithm}
\usepackage{graphics}
\usepackage{multirow}
\input{preamble}

\definecolor{cvprblue}{rgb}{0.21,0.49,0.74}
\usepackage[pagebackref,breaklinks,colorlinks,citecolor=cvprblue]{hyperref}

\input{macros}

\def\paperID{16812} % *** Enter the Paper ID here
\def\confName{CVPR}
\def\confYear{2024}

\title{EMCAD: Efficient Multi-scale Convolutional Attention Decoding for Medical Image Segmentation}

\author{Md Mostafijur Rahman, Mustafa Munir, and Radu Marculescu\\
The University of Texas at Austin\\
Austin, Texas, USA\\
{\tt\small {mostafijur.rahman, mmunir, radum}@utexas.edu}
}

\begin{document}
\maketitle
\input{sec/0_abstract}    
\input{sec/1_intro}
\input{sec/2_related_works}
\input{sec/3_methodology}
\input{sec/4_experiments}

\input{sec/5_ablation_study}

\input{sec/6_conclusion}
{
    \small
    \bibliographystyle{ieeenat_fullname}
    \bibliography{main}
}
% WARNING: do not forget to delete the supplementary pages from your submission 
\input{sec/X_suppl}

\end{document}

%% file: preamble.tex
\usepackage[dvipsnames]{xcolor}

%% file: macros.tex
\usepackage{times}
\usepackage{multirow}
\usepackage{graphicx}
\usepackage[utf8]{inputenc}
\usepackage{amsmath}
\usepackage{enumitem}
\usepackage{amsfonts}
\usepackage{algorithm}
\usepackage{algpseudocode}
\usepackage{comment}
\usepackage{multirow}
\usepackage{listings}
\usepackage{xcolor}

\definecolor{codegreen}{rgb}{0,0.6,0}
\definecolor{codegray}{rgb}{0.5,0.5,0.5}
\definecolor{codepurple}{rgb}{0.58,0,0.82}
\definecolor{backcolour}{rgb}{0.95,0.95,0.92}

\lstdefinestyle{mystyle}{
    backgroundcolor=\color{backcolour},   
    commentstyle=\color{codegreen},
    keywordstyle=\color{magenta},
    numberstyle=\tiny\color{codegray},
    stringstyle=\color{codepurple},
    basicstyle=\ttfamily\footnotesize,
    breakatwhitespace=false,         
    breaklines=true,                 
    captionpos=b,                    
    keepspaces=true,                 
    numbers=left,                    
    numbersep=5pt,                  
    showspaces=false,                
    showstringspaces=false,
    showtabs=false,                  
    tabsize=2
}

%% file: sec/0_abstract.tex
\begin{abstract}
%Deep learning has shown great success in computer vision. However, it has generally huge computational and infrastructural costs due to depending on large training datasets.
%U-shaped networks having pretrained transformers and CNN encoders are widely used for medical image segmentation. 

An efficient and effective decoding mechanism is crucial in medical image segmentation, especially in scenarios with limited computational resources. However, these decoding mechanisms usually come with high computational costs. To address this concern, we introduce EMCAD, a new efficient multi-scale convolutional attention decoder, designed to optimize both performance and computational efficiency. EMCAD leverages a unique multi-scale depth-wise convolution block, significantly enhancing feature maps through multi-scale convolutions. EMCAD also employs channel, spatial, and grouped (large-kernel) gated attention mechanisms, which are highly effective at capturing intricate spatial relationships while focusing on salient regions. By employing group and depth-wise convolution, EMCAD is very efficient and scales well (e.g., only 1.91M parameters and 0.381G FLOPs are needed when using a standard encoder). Our rigorous evaluations across 12 datasets that belong to six medical image segmentation tasks reveal that EMCAD achieves state-of-the-art (SOTA) performance with 79.4\% and 80.3\% reduction in \#Params and \#FLOPs, respectively. Moreover, EMCAD’s adaptability to different encoders and versatility across segmentation tasks further establish EMCAD as a promising tool, advancing the field towards more efficient and accurate medical image analysis. Our implementation is available at https://github.com/SLDGroup/EMCAD. %\href{https://github.com/SLDGroup/EMCAD}{EMCAD}.

%This is further enriched by incorporating channel, spatial, and grouped gated attention mechanisms, which are highly effective in discerning intricate spatial relationships and concentrating on key regions.

%In U-shaped networks, the hierarchical features extracted from the transformers and CNN encoders are processed using expanding networks/decoders to get high-resolution segmentation outputs. However, these decoders usually come with high computational costs. To this end, we introduce Efficient Multi-scale Convolutional Attention Decoding (EMCAD), a fully convolutional approach specifically designed for 2D medical image segmentation. EMCAD integrates efficient multi-scale convolutions and attention mechanisms, adeptly refining feature maps from hierarchical encoders. The multi-scale feature extraction with attention mechanism facilitates the capture of intricate details in medical images, ensuring high segmentation accuracy. Rigorous evaluations across several medical image segmentation benchmarks reveal that EMCAD achieves state-of-the-art performance with 75.8\% and 81.46\% reduction in \#Parameters and \#FLOPs, respectively. Moreover, EMCAD’s adaptability to different encoders and versatility across segmentation tasks further establish EMCAD as a potent tool, advancing the field towards more efficient and accurate medical image analysis.}
%, making it a practical choice for real-world applications
\end{abstract}
\vspace{-.2cm}

%% file: sec/1_intro.tex
\section{Introduction}
\label{sec:intro}

In the realm of medical diagnostics and therapeutic strategies, automated segmentation of medical images is vital, as it classifies pixels to identify critical regions such as lesions, tumors, or entire organs. A variety of U-shaped convolutional neural network (CNN) architectures \cite{ronneberger2015u, oktay2018attention, zhou2018unet++, huang2020unet, fan2020pranet, lou2021dc}, notably UNet \cite{ronneberger2015u}, UNet++ \cite{zhou2018unet++}, UNet3+ \cite{huang2020unet}, and nnU-Net \cite{isensee2021nnu}, have become standard techniques for this purpose, achieving high-quality, high-resolution segmentation output. Attention mechanisms \cite{oktay2018attention, chen2018reverse, fan2020pranet, woo2018cbam, dong2021polyp} have also been integrated into these models to enhance feature maps and improve pixel-level classification. Although attention-based models have shown improved performance, they still face significant challenges due to the computationally expensive convolutional blocks that are typically used in conjunction with attention mechanisms.
% 

%Recently, vision transformers \cite{dosovitskiy2020image} have shown great promise for capturing long-range dependencies among pixels and demonstrated improved performance, particularly for medical image segmentation \cite{chen2021transunet, cao2021swin, dong2021polyp, wang2022stepwise, Rahman_2023_WACV, rahman2023multi, zhang2021transfuse, wang2022uctransnet}. The self-attention (SA) mechanism used in transformers learns correlations among input patches; this enables capturing the long-range dependencies among pixels. Recently, hierarchical vision transformers such as the Swin transformer \cite{liu2021swin}, the pyramid vision transformer (PVT) \cite{wang2021pyramid},  MaxViT \cite{tu2022maxvit}, MERIT \cite{rahman2023multi}, have been introduced to enhance performance. These hierarchical vision transformers are effective in medical image segmentation tasks \cite{chen2021transunet, cao2021swin, dong2021polyp, wang2022stepwise, Rahman_2023_WACV, rahman2023multi}. As self-attention modules employed in transformers have limited capacity to learn (local) spatial relationships among pixels \cite{chu2021conditional, islam2020much}, some methods \cite{xie2021segformer, wang2022uformer, wang2022pvt, dong2021polyp, wang2022stepwise, Rahman_2023_WACV, rahman2023multi} incorporate local convolutional attention modules in the decoder. However, due to the locality of convolution operations, these methods have difficulties at capturing long-range correlations among pixels.

Recently, vision transformers \cite{dosovitskiy2020image} have shown promise in medical image segmentation tasks \cite{chen2021transunet, cao2021swin, dong2021polyp, wang2022stepwise, Rahman_2023_WACV, rahman2023multi, zhang2021transfuse, wang2022uctransnet} by capturing long-range dependencies among pixels through Self-attention (SA) mechanisms. Hierarchical vision transformers like Swin \cite{liu2021swin}, PVT \cite{wang2021pyramid, wang2022pvt}, MaxViT \cite{tu2022maxvit}, MERIT \cite{rahman2023multi}, ConvFormer \cite{lin2023convformer}, and MetaFormer \cite{yu2022metaformer} have been introduced to further improve the performance in this field. While the SA excels at capturing global information, it is less adept at understanding the local spatial context \cite{chu2021conditional, islam2020much}. To address this limitation, some approaches have integrated local convolutional attention within the decoders to better grasp spatial details. Nevertheless, these methods can still be computationally demanding because they frequently employ costly convolutional blocks. This limits their applicability to real-world scenarios where computational resources are restricted.

%Despite their success, these methods can still be computationally demanding because they frequently employ costly convolutional blocks.

\begin{figure}[t]
\begin{center}
%\fbox{\rule{0pt}{2in} \rule{.9\linewidth}{0pt}}
\includegraphics[width=0.85\linewidth]{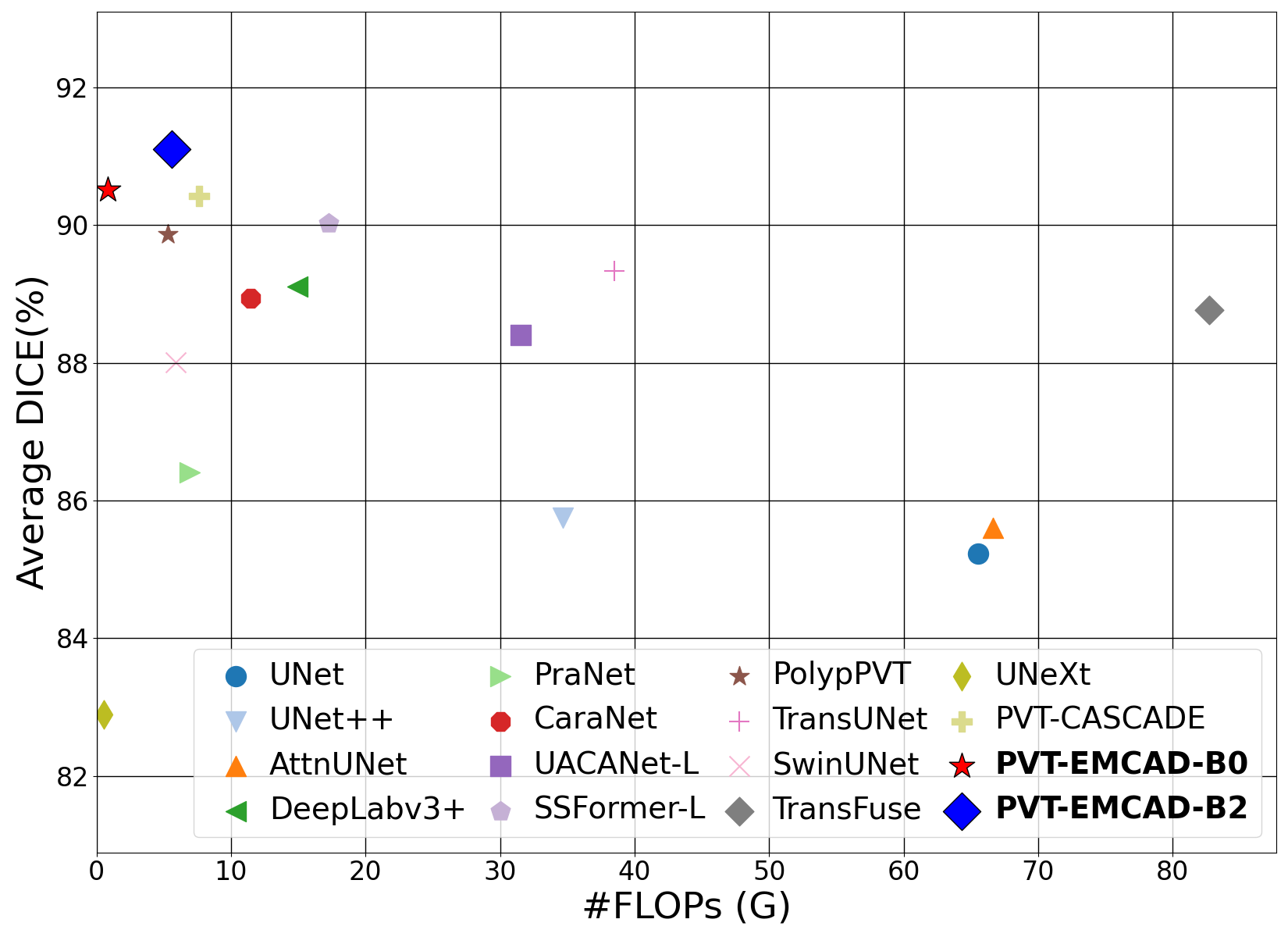}
\end{center}
\vspace{-0.6cm}
   \caption{Average DICE scores vs. \#FLOPs for different methods over 10 binary medical image segmentation datasets. As shown, our approaches (PVT-EMCAD-B0 and PVT-EMCAD-B2) have the lowest \#FLOPs, yet the highest DICE scores.}
   \vspace{-0.4cm}
\label{fig:dice_vs_flops}
\end{figure}

To address the aforementioned limitations, we introduce EMCAD, an efficient multi-scale convolutional attention decoding using a new multi-scale depth-wise convolution block. More precisely, EMCAD enhances the feature maps via efficient multi-scale convolutions, while incorporating complex spatial relationships and local attention through the use of channel, spatial, and grouped (large-kernel) gated attention mechanisms. Our contributions are as follows:
%CASCADE consists of attention gates (AGs) to fuse features with skip connections and convolutional attention modules (CAMs) to refine features. 
\begin{itemize}
  \item \textbf{New Efficient Multi-scale Convolutional Decoder:} We introduce an efficient multi-scale cascaded fully-convolutional attention decoder (EMCAD) for 2D medical image segmentation; this takes the multi-stage features of vision encoders and progressively enhances the multi-scale and multi-resolution spatial representations. EMCAD has only 0.506M parameters and 0.11G FLOPs for a tiny encoder with \#channels = [32, 64, 160, 256], while it has 1.91M parameters and 0.381G FLOPs for a standard encoder with \#channels = [64, 128, 320, 512].
\item \textbf{Efficient Multi-scale Convolutional Attention Module:} We introduce MSCAM, a new efficient multi-scale convolutional attention module that performs depth-wise convolutions at multiple scales; this refines the feature maps produced by vision encoders and enables capturing multi-scale salient features by suppressing irrelevant regions. The use of depth-wise convolutions makes MSCAM very efficient. 
\item \textbf{Large-kernel Grouped Attention Gate:} We introduce a new grouped attention gate to fuse refined features with the features from skip connections. By using larger kernel ($3\times3$) group convolutions instead of point-wise convolutions in the design, we capture salient features in a larger local context with less computation.
\item \textbf{Improved Performance:} We empirically show that EMCAD can be used with \textit{any} hierarchical vision encoder (e.g., PVTv2-B0, PVTv2-B2 \cite{wang2022pvt}), while significantly improving the performance of 2D medical image segmentation. EMCAD produces better results than SOTA methods with a significantly lower computational cost (as shown in Figure \ref{fig:dice_vs_flops}) on 12 medical image segmentation benchmarks that belong to six different tasks.
\end{itemize}

The remaining of this paper is organized as follows: Section \ref{sec:related_work} summarizes related work. Section \ref{sec:methodology} describes the proposed method. Section \ref{sec:experiments} explains our experimental setup and results on 12 medical image segmentation benchmarks. Section \ref{sec:ablation_study} covers different ablation experiments. Lastly, Section \ref{sec:conclusion} concludes the paper. 
%-------------------------------------------------------------------------

%\vspace{-.2cm}

%% file: sec/2_related_works.tex
\section{Related Work}
\label{sec:related_work}
%We divide the related work into three parts, i.e., vision transformers, vision graph convolutional networks, and medical image segmentation; these are described next. 

\subsection{Vision encoders}
Convolutional Neural Networks (CNNs) \cite{krizhevsky2012imagenet, simonyan2014very, he2016deep, szegedy2015going, howard2017mobilenets, sandler2018mobilenetv2, hu2018squeeze, tan2019efficientnet, liu2022convnet} have been foundational as encoders due to their proficiency in handling spatial relationships in images. More precisely, AlexNet \cite{krizhevsky2012imagenet} and VGG \cite{simonyan2014very} pave the way, leveraging deep layers of convolutions to extract features progressively. GoogleNet \cite{szegedy2015going} introduces the inception module, allowing more efficient computation of representations across various scales. ResNet \cite{he2016deep} introduces residual connections, enabling the training of networks with substantially more layers by addressing the vanishing gradients problem. MobileNets \cite{howard2017mobilenets, sandler2018mobilenetv2} bring CNNs to mobile devices through lightweight, depth-wise separable convolutions. EfficientNet \cite{tan2019efficientnet} introduces a scalable architectural design to CNNs with compound scaling. Although CNNs are pivotal for many vision applications, they generally lack the ability to capture long-range dependencies within images due to their inherent local receptive fields.

 Recently, Vision Transformers (ViTs), pioneered by Dosovitskiy et al. \cite{dosovitskiy2020image}, enabled the learning of long-range relationships among pixels using Self-attention (SA). Since then, ViTs have been enhanced by integrating CNN features \cite{wang2022pvt,tu2022maxvit}, developing novel self-attention (SA) blocks \cite{liu2021swin,tu2022maxvit}, and introducing new architectural designs \cite{wang2021pyramid, xie2021segformer}. The Swin Transformer \cite{liu2021swin} incorporates a sliding window attention mechanism, while SegFormer \cite{xie2021segformer} leverages Mix-FFN blocks for hierarchical structures. PVT \cite{wang2021pyramid} uses spatial reduction attention, refined in PVTv2 \cite{wang2022pvt} with overlapping patch embedding and a linear complexity attention layer. MaxViT \cite{tu2022maxvit} introduces a multi-axis self-attention to form a hierarchical CNN-transformer encoder. Although ViTs address the CNNs limitation in capturing long-range pixel dependencies \cite{krizhevsky2012imagenet, simonyan2014very, he2016deep, szegedy2015going, howard2017mobilenets, sandler2018mobilenetv2, hu2018squeeze, tan2019efficientnet, liu2022convnet}, they face challenges in capturing the local spatial relationships among pixels. In this paper, we aim to overcome these limitations by introducing a new multi-scale cascaded attention decoder that refines feature maps and incorporates local attention using a multi-scale convolutional attention module.

\vspace{-.1cm}
\subsection{Medical image segmentation}

%Medical image segmentation is the task of classifying pixels into lesions, tumours, or organs in a medical image (e.g., endoscopy, MRI, and CT) \cite{chen2021transunet}. To address this task, U-shaped architectures \cite{ronneberger2015u, oktay2018attention, zhou2018unet++, huang2020unet, lou2021dc} have been commonly utilized due to their sophisticated encoder-decoder structure. Ronneberger et al. \cite{huang2020unet} introduce UNet, an encoder-decoder architecture that utilizes skip connections to aggregate features from multiple stages. In UNet++ \cite{zhou2018unet++}, nested encoder-decoder sub-networks are connected through dense skip connections. UNet 3+ \cite{huang2020unet} further extends this concept by exploring full-scale skip connections with intra-connections among the decoder blocks. DC-UNet \cite{lou2021dc} incorporates the multi-resolution convolution block and residual path within skip connections. These architectures have proven to be effective in medical image segmentation tasks.

Medical image segmentation involves pixel-wise classification to identify various anatomical structures like lesions, tumors, or organs within different imaging modalities such as endoscopy, MRI, or CT scans \cite{chen2021transunet}. U-shaped networks \cite{ronneberger2015u, oktay2018attention, zhou2018unet++, huang2020unet, lou2021dc, ibtehaz2023acc, chen2022aau, isensee2021nnu} are particularly favored due to their simple but effective encoder-decoder design. The UNet \cite{ronneberger2015u} pioneered this approach with its use of skip connections to fuse features at different resolution stages. UNet++ \cite{zhou2018unet++} evolves this design by incorporating nested encoder-decoder pathways with dense skip connections. Expanding on these ideas, UNet 3+ \cite{huang2020unet} introduces comprehensive skip pathways that facilitate full-scale feature integration. Further advancement comes with DC-UNet \cite{lou2021dc}, which integrates a multi-resolution convolution scheme and residual paths into its skip connections. The DeepLab series, including DeepLabv3 \cite{chen2017deeplab} and DeepLabv3+ \cite{chen2018encoder}, introduce atrous convolutions and spatial pyramid pooling to handle multi-scale information. SegNet \cite{badrinarayanan2017segnet} uses pooling indices to upsample feature maps, preserving the boundary details. nnU-Net \cite{isensee2021nnu} automatically configures hyperparameters based on the specific dataset characteristics, using standard 2D and 3D UNets. Collectively, these U-shaped models have become a benchmark for success in the domain of medical image segmentation.

Recently, vision transformers have emerged as a formidable force in medical image segmentation, harnessing the ability to capture pixel relationships at global scales \cite{cao2021swin, chen2021transunet, dong2021polyp, Rahman_2023_WACV, rahman2023multi, wang2022uctransnet, zhang2021transfuse, xie2021segformer}. TransUNet \cite{chen2021transunet} presents a novel blend of CNNs for local feature extraction and transformers for global context, enhancing both local and global feature capture. Swin-Unet \cite{cao2021swin} extends this by incorporating Swin Transformer blocks \cite{liu2021swin} into a U-shaped model for both encoding and decoding processes. Building on these concepts, MERIT \cite{rahman2023multi} introduces a multi-scale hierarchical transformer, which employs SA across different window sizes, thus enhancing the model capacity to capture multi-scale features critical for medical image segmentation.

The integration of attention mechanisms has been investigated within CNNs \cite{oktay2018attention, fan2020pranet} and transformer-based systems \cite{dong2021polyp} for enhancing medical image segmentation. PraNet \cite{fan2020pranet} employs a reverse attention strategy for feature refinement. PolypPVT \cite{dong2021polyp} leverages PVTv2 \cite{wang2022pvt} as its backbone encoder and incorporates CBAM \cite{woo2018cbam} within its decoding stages. The CASCADE \cite{Rahman_2023_WACV} presents a novel cascaded decoder, combining channel \cite{hu2018squeeze} and spatial \cite{chen2017sca} attention to refine features at multiple stages, extracted from a transformer encoder, culminating in high-resolution segmentation outputs. While CASCADE achieves notable performance in segmenting medical images by integrating local and global insights from transformers, it is computationally inefficient due to the use of triple $3\times3$ convolution layers at each decoder stage. In addition to this, it uses single-scale convolutions during decoding. Our new proposal involves the adoption of multi-scale depth-wise convolutions to mitigate these constraints.

%Attention mechanisms have also been explored in combination with both CNNs \cite{oktay2018attention, fan2020pranet} and transformer-based architectures \cite{dong2021polyp} in medical image segmentation. PraNet \cite{fan2020pranet} utilizes the reverse attention mechanism \cite{chen2018reverse}. In PolypPVT \cite{dong2021polyp}, authors employ PVTv2 \cite{wang2022pvt} as the encoder and integrates CBAM \cite{woo2018cbam} attention blocks in the decoder, along with other modules. CASCADE \cite{Rahman_2023_WACV} proposes a cascaded decoder that utilizes both channel attention \cite{hu2018squeeze} and spatial attention \cite{chen2017sca} modules for feature refinement. CASCADE extracts features from four stages of the transformer encoder and uses cascaded refinement to generate high-resolution segmentation maps.
%Due to incorporating local information with global information of transformers, CASCADE exhibits remarkable performance in medical image segmentation. However, CASCADE decoder has two main limitations: i) long-range attention deficit due to using only convolution operations during decoding and ii) high computational inefficiency due to using three $3\times3$ convolutions in each stage of the decoder. We propose to use graph convolution to overcome these limitations. 

% \begin{figure*}[htb]

% \begin{minipage}[b]{1.0\linewidth}
%   \centering
%   \centerline{\includegraphics[width=16cm]{images/Architecture.png}}
% %  \centerline{(a) Result 1}\medskip
% \end{minipage}

% \caption{Network architecture.}
% \label{fig:architecture}
% %
% \end{figure*}

\vspace{-.2cm}

%% file: sec/3_methodology.tex
\section{Methodology}
\label{sec:methodology}

In this section, we first introduce our new EMCAD decoder and then explain two transformer-based architectures (i.e., PVT-EMCAD-B0 and PVT-EMCAD-B2) incorporating our proposed decoder.

\begin{figure*}[t]
\begin{center}
%\fbox{\rule{0pt}{2in} \rule{.9\linewidth}{0pt}}
\includegraphics[width=0.9\linewidth]{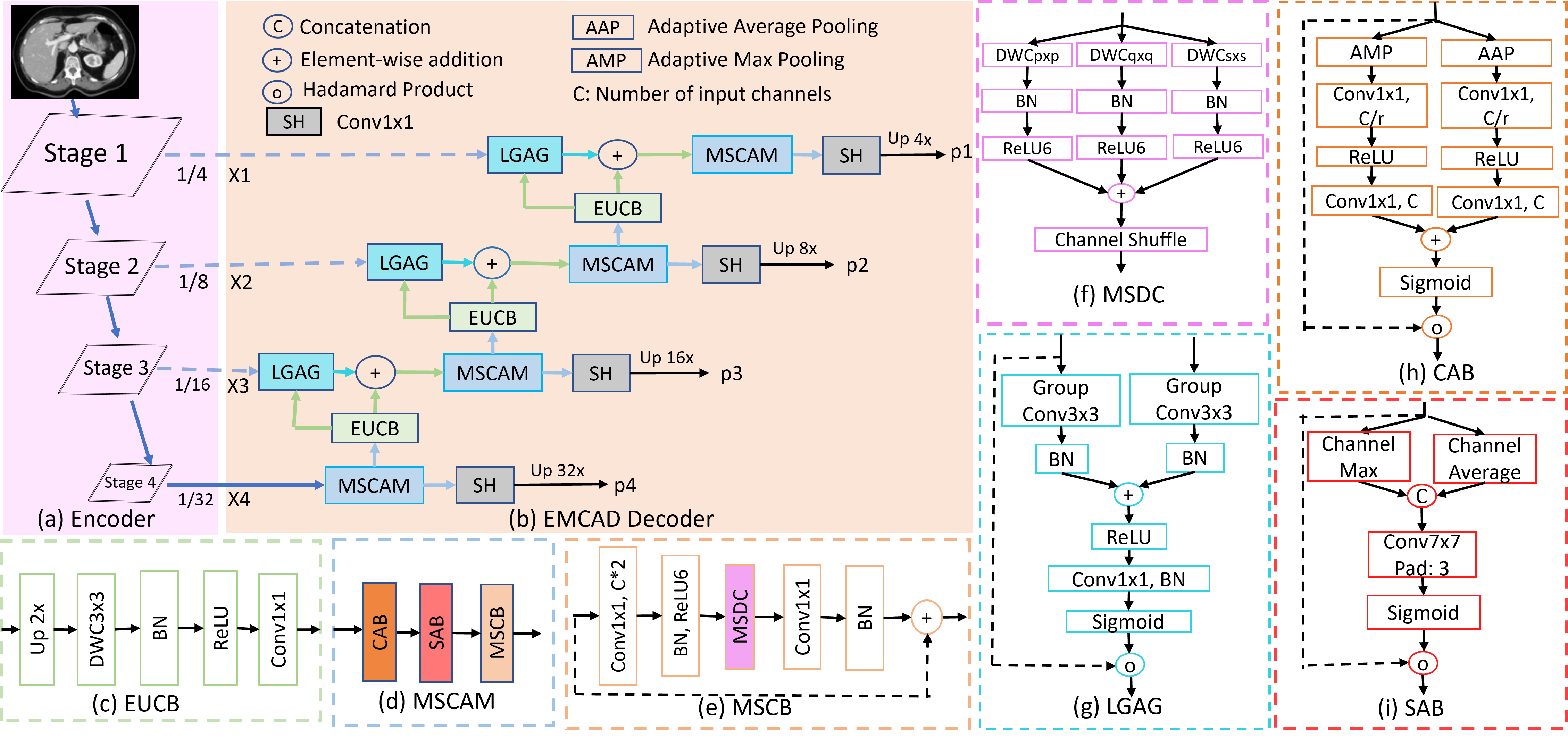}
\end{center}
\vspace{-.6cm}
   \caption{Hierarchical encoder with newly proposed EMCAD decoder architecture. (a) CNN or transformer encoder with four hierarchical stages, (b) EMCAD decoder, (c) Efficient up-convolution block (EUCB), (d) Multi-scale convolutional attention module (MSCAM), (e) Multi-scale convolution block (MSCB), (f) Multi-scale (parallel) depth-wise convolution (MSDC), (g) Large-kernel grouped attention gate (LGAG), (h) Channel attention block (CAB), and (i) Spatial attention block (SAB). X1, X2, X3, and X4 are the features from the four stages of the hierarchical encoder. p1, p2, p3, and p4 are output segmentation maps from four stages of our decoder.}
   \vspace{-.5cm}
\label{fig:architecture}
\end{figure*}

\vspace{-.1cm}
\subsection{Efficient multi-scale convolutional attention decoding (EMCAD)}
%Existing transformer-based models have limited (local) contextual information processing ability among pixels. As a result, the transformer-based model faces difficulties in locating the more discriminating local features. To address this issue, some works \cite{dong2021polyp,Rahman_2023_WACV, rahman2023multi} utilize computationally expensive 2D convolution blocks in the decoder. Although the convolution block helps to incorporate the local information, it results in long-range attention deficits. To overcome this problem, we propose a new cascaded graph convolutional decoder, G-CASCADE, for pyramid encoders.
In this section, we introduce our efficient multi-scale convolutional decoding (EMCAD) to process the multi-stage features extracted from pretrained hierarchical vision encoders for high-resolution semantic segmentation. As shown in Figure \ref{fig:architecture}(b), EMCAD consits of efficient multi-scale convolutional attention modules (MSCAMs) to robustly enhance the feature maps, large-kernel grouped attention gates (LGAGs) to refine feature maps fusing with the skip connection via gated attention mechanism, efficient up-convolution blocks (EUCBs) for up-sampling followed by enhancement of feature maps, and segmentation heads (SHs) to produce the segmentation outputs. 

More specifically, we use four MSCAMs to refine pyramid features (i.e., $X1$, $X2$, $X3$, $X4$ in Figure \ref{fig:architecture}) extracted from the four stages of the encoder. After each MSCAM, we use an SH to produce a segmentation map of that stage. Subsequently, we upscale the refined feature maps using EUCBs and add them to the outputs from the corresponding LGAGs. Finally, we add four different segmentation maps to produce the final segmentation output. Different modules of our decoder are described next.

%The experimental section compares CASCADE with other existing decoders with various encoders that can generate feature pyramids. We compare the attention distribution before the final prediction of several typical multi-stage feature aggregation decoders for Transformers. As demonstrated in Figure 2(a), after CASCADE fuses multi-stage features, the prediction head can accurately focus on critical targets. In addition, our CASCADE can be used for other Pyramid Transformer encoders and can improve the model’s accuracy. There is a further demonstration in Section 3.3.
\vspace{-.2cm}
\subsubsection{Large-kernel grouped attention gate (LGAG)}
%emphasizing salient features and suppressing irrelevant ones
We introduce a new \textit{large-kernel grouped attention gate} (LGAG) to progressively combine feature maps with attention coefficients, which are learned by the network to allow higher activation of relevant features and suppression of irrelevant ones. This process employs a gating signal derived from higher-level features to control the flow of information across different stages of the network, thus enhancing its precision for medical image segmentation. Unlike Attention UNet \cite{oktay2018attention} which uses $1\times1$ convolution to process gating signal $g$ (features from skip connections) and input feature map $x$ (upsampled features), in our $q_{att}$(.) function, we process $g$ and $x$ by applying separate $3\times3$ group convolutions $GC_g$(.) and $GC_x$(.), respectively. These convolved features are then normalized using batch normalization ($BN$(.)) \cite{ioffe2015batch} and merged through element-wise addition. The resultant feature map is activated through a ReLU ($R$(.)) layer \cite{nair2010rectified}. Afterward, we apply a $1\times1$ convolution ($C$(.)) followed by $BN$(.) layer to get a single channel feature map. We then pass the resultant single-channel feature map through a Sigmoid ($\sigma$(.)) activation function to yield the attention coefficients. The output of this transformation is used to scale the input feature $x$ through element-wise multiplication, producing the attention-gated feature $LGAG(g,x)$. The $LGAG$(·) (Figure \ref{fig:architecture}(g)) can be formulated as in Equations \ref{eq:ags_1} and \ref{eq:ags_2}: \vspace{-.1cm}
\begin{equation}
    q_{att}(g, x) = R(BN(GC_g(g) + BN(GC_x(x)))))
    \vspace{-.1cm}
    \label{eq:ags_1}
\end{equation}
\vspace{-.1cm}
\begin{equation}
   LGAG(g, x) = x \circledast \sigma(BN(C(q_{att}(g, x))))
   \vspace{-.1cm}
    \label{eq:ags_2}
\end{equation} 
Due to using $3\times3$ kernel group convolutions in $q_{att}$(.), our LGAG captures comparatively larger spatial contexts with less computational cost.

\vspace{-.2cm}
\subsubsection{Multi-scale convolutional attention module (MSCAM)}
We introduce an efficient multi-scale convolutional attention module to refine the feature maps. MSCAM consists of a channel attention block ($CAB$(·)) to put emphasis on pertinent channels, a spatial attention block \cite{chen2017sca} ($SAB$(·)) to capture the local contextual information, and an efficient multi-scale convolution block ($MSCB$(.)) to enhance the feature maps preserving contextual relationships. The $MSCAM$(.) (Figure \ref{fig:architecture}(d)) is given in Equation \ref{eq:mscam}: \vspace{-.1cm}
\begin{equation}
    MSCAM(x) = MSCB(SAB(CAB(x)))
    \vspace{-.1cm}
    \label{eq:mscam}
\end{equation}
where $x$ is the input tensor. Due to using depth-wise convolution in multiple scales, our MSCAM is more effective with significantly lower computational cost than the convolutional attention module (CAM) proposed in \cite{Rahman_2023_WACV}.

\textbf{Multi-scale Convolution Block (MSCB):} We introduce an efficient multi-scale convolution block to enhance the features generated by our cascaded expanding path. In our MSCB, we follow the design of the inverted residual block (IRB) of MobileNetV2 \cite{sandler2018mobilenetv2}. However, unlike IRB, our MSCB performs depth-wise convolution at multiple scales and uses channel\_shuffle \cite{zhang2018shufflenet} to shuffle channels across groups. More specifically, in our MSCB, we first expand the number of channels (i.e., expansion\_factor = 2) using a point-wise ($1 \times 1$) convolution layers $PWC_1$(·) followed by a batch normalization layer $BN$(·) and a ReLU6 \cite{krizhevsky2010convolutional} activation layer $R6$(.). We then use a multi-scale depth-wise convolution $MSDC$(.) to capture both multi-scale and multi-resolution contexts. As depth-wise convolution overlooks the relationships among channels, we use a channel\_shuffle operation to incorporate relationships among channels. Afterward, we use another point-wise convolution $PWC_2$(.) followed by a $BN$(.) to transform back the original \#channels, which also encodes dependency among channels. The $MSCB$(·) (Figure \ref{fig:architecture}(e)) is formulated as in Equation \ref{eq:mscb}: \vspace{-.1cm}
%\scriptsize
\begin{equation}
\scalebox{0.76}{\(
    MSCB(x) = BN(PWC_2(CS(MSDC(R6(BN(PWC_1(x)))))))
\)}
\vspace{-.1cm}
    \label{eq:mscb}
\end{equation}
%\normalsize
where parallel $MSDC$(.) (Figure \ref{fig:architecture}(f)) for different kernel sizes ($KS$) can be formulated using Equation \ref{eq:msdc}: \vspace{-.1cm}
\begin{equation}
    \scalebox{0.95}{\( 
    MSDC(x) = \sum_{ks \in KS} DWCB_{ks}(x)
    \)}
    \vspace{-.1cm}
    \label{eq:msdc}
\end{equation}
where $DWCB_{ks}(x) = R6(BN(DWC_{ks}(x)))$. Here, $DWC_{ks}$(.) is a depth-wise convolution with the kernel size $ks$. $BN$(.) and $R6$(.) are batch normalization and ReLU6 activation, respectively. Additionally, our sequential $MSDC$(.) uses the recursively updated input $x$, where the input $x$ is residually connected to the previous $DWCB_{ks}$(.) for better regularization as in Equation \ref{eq:msdc_sequential}: \vspace{-.1cm}
\begin{equation}
    x = x + DWCB_{ks}(x)
    \vspace{-.1cm}
    \label{eq:msdc_sequential}
\end{equation}

% need to cite BN ReLU6, and ReLU paper
\textbf{Channel Attention Block (CAB):} We use channel attention block to assign different levels of importance to each channel, thus emphasizing more relevant features while suppressing less useful ones. Basically, the CAB identifies \textit{which feature maps} to focus on (and then refine them). Following \cite{woo2018cbam}, in CAB, we first apply the adaptive maximum pooling ($P_m$(·)) and adaptive average pooling ($P_a$(·)) to the spatial dimensions (i.e., height and width) to extract the most significant feature of the entire feature map per channel. Then, for each pooled feature map, we reduce the number of channels $r=1/16$ times separately using a point-wise convolution ($C_1$(·)) followed by a ReLU activation ($R$). Afterward, we recover the original channels using another point-wise convolution ($C_2$(·)). We then add both recovered feature maps and apply Sigmoid ($\sigma$) activation to estimate attention weights. Finally, we incorporate these weights to input $x$ using the Hadamard product ($\circledast$). The $CAB$(·) (Figure \ref{fig:architecture}(h)) is defined using Equation \ref{eq:chab}: \vspace{-.1cm}
 %\small 
 \begin{equation}
 \scalebox{0.8}{\( CAB(x)=\sigma(C_2(R(C_1(P_m(x))))+C_2(R(C_1(P_a(x))))) \circledast x
 \)}
 \vspace{-.1cm}
    \label{eq:chab}
\end{equation}
%\normalsize
%where $\circledast$ is the Hadamard product. 
%$\sigma$(·) is the Sigmoid activation. $P_m$(·) and $P_a$(·) denote adaptive maximum pooling and adaptive average pooling, respectively. $C_1$(·) is a convolutional layer with $1 \times 1$ kernel size to reduce the channel dimension 16 times. $R$ is a ReLU activation layer and $C_2$(·) is another convolutional layer to recover the original channel dimension. 

\textbf{Spatial Attention Block (SAB):} We use spatial attention to mimic the attentional processes of the human brain by focusing on specific parts of an input image. Basically, the SAB determines \textit{where} to focus in a feature map; then it enhances those features. This process enhances the model's ability to recognize and respond to relevant spatial features, which is crucial for image segmentation where the context and location of objects significantly influence the output. In SAB, we first pool maximum ($Ch_{max}$(·)) and average ($Ch_{avg}$(·)) values along the channel dimension to pay attention to local features. Then, we use a large kernel (i.e., $7 \times 7$ as in \cite{dong2021polyp}) convolution layer to enhance local contextual relationships among features. Afterward, we apply the Sigmoid activation ($\sigma$) to calculate attention weights. Finally, we feed these weights to the input $x$ (using Hadamard product ($\circledast$) to attend information in a more targeted way. The $SAB$(.) (Figure \ref{fig:architecture}(i)) is defined using Equation \ref{eq:spab}:
\small
\begin{equation}
    SAB(x) = \sigma(LKC([Ch_{max}(x), Ch_{avg}(x)])) \circledast x
    \vspace{-0.1cm}
    \label{eq:spab}
\end{equation}
\normalsize
 
\vspace{-.4cm}
\subsubsection{Efficient up-convolution block (EUCB)}
We use an efficient up-convolution block to progressively upsample the feature maps of the current stage to match the dimension and resolution of the feature maps from the next skip connection. The EUCB first uses an UpSampling $Up$(·) with scale-factor 2 to upscale the feature maps. Then, it enhances the upscaled feature maps by applying a $3\times3$ depth-wise convolution $DWC$(·) followed by a $BN$(·) and a $ReLU$(.) activation. Finally, a $1\times1$ convolution $C_{1\times1}$(.) is used to reduce the \#channels to match with the next stage. The $EUCB$(·) (Figure \ref{fig:architecture}(c)) is formulated as in Equation \ref{eq:up_conv}:
\begin{equation}
    EUCB(x) = C_{1\times1}(ReLU(BN(DWC(Up(x)))))
    \vspace{-0.1cm}
    \label{eq:up_conv}
\end{equation}
Due to using depth-wise convolution instead of $3\times3$ convolution, our EUCB is \textit{very efficient}.  
\vspace{-.3cm}
\subsubsection{Segmentation head (SH)}
We use segmentation heads to produce the segmentation outputs from the refined feature maps of four stages of the decoder. The SH layer applies a $1\times1$ convolution $Conv_{1\times1}$(·) to the refined feature maps having $ch_i$ channels ($ch_i$ is the \#channels in the feature map of stage $i$) and produces output with \#channels equal to \#classes in target dataset for multi-class but $1$ channel for binary segmentation. The $SH$(·) is formulated as in Equation \ref{eq:seg_head}:
\begin{equation}
    SH(x) = Conv_{1\times1}(x)
    \vspace{-0.1cm}
    \label{eq:seg_head}
\end{equation}
\vspace{-.4cm}
\subsection{Overall architecture}
To show the generalization, effectiveness, and ability to process multi-scale features for medical image segmentation, we integrate our EMCAD decoder alongside tiny (PVTv2-B0) and standard (PVTv2-B2) networks of PVTv2 \cite{wang2022pvt}. However, our decoder is adaptable and seamlessly compatible with other hierarchical backbone networks.

%To ensure effective generalization and the ability to process multi-scale features in medical image segmentation, we integrate our proposed EMCAD decoder with two different hierarchical backbone encoder networks such as PVTv2 \cite{wang2022pvt} and MERIT \cite{rahman2023multi}. PVTv2 utilizes convolution operations instead of traditional transformer patch embedding modules to consistently capture the spatial information. MERIT utilizes two MaxViT \cite{tu2022maxvit} encoders with varying window sizes for self-attention, thus enabling it to capture multi-scale features. 
%\textbf{PVT-EMCAD:} 
PVTv2 differs from conventional transformer patch embedding modules by applying convolutional operations for consistent spatial information capture.
Using PVTv2-b0 (Tiny) and PVTv2-b2 (Standard) encoders \cite{wang2022pvt}, we develop the PVT-EMCAD-B0 and PVT-EMCAD-B2 architectures. To adopt PVTv2, we first extract the features (X1, X2, X3, and X4) from four layers and feed them (i.e., X4 in the upsample path and X3, X2, X1 in the skip connections) into our EMCAD decoder as shown in Figure \ref{fig:architecture}(a-b). EMCAD then processes them and produces four segmentation maps that correspond to the four stages of the encoder network.

\vspace{-.1cm}
\subsection{Multi-stage loss and outputs aggregation}
%We get four output segmentation maps $p_1$, $p_2$, $p_3$, and $p_4$ from the four prediction heads for the four stages of our EMCAD decoder. 
Our EMCAD decoder's four segmentation heads produce four prediction maps $p_1$, $p_2$, $p_3$, and $p_4$ across its stages.

\textbf{Loss aggregation:} We adopt a combinatorial approach to loss combination called MUTATION, inspired by the work of MERIT \cite{rahman2023multi} for multi-class segmentation. This involves calculating the loss for all possible combinations of predictions derived from 
$4$ heads, totaling $2^4-1=15$ unique predictions, and then summing these losses. We focus on minimizing this cumulative combinatorial loss during the training process. For binary segmentation, we optimize the additive loss like \cite{Rahman_2023_WACV} with an additional term $\mathcal{L}_{p_1 + p_2 + p_3 + p_4}$ as in Equation \ref{eq:add_loss}:
\begin{equation}
\scalebox{0.85}{\(
   \mathcal{L}_{total} = \alpha\mathcal{L}_{p_1} + \beta\mathcal{L}_{p_2} + \gamma\mathcal{L}_{p_3} + \zeta\mathcal{L}_{p_4} + \delta\mathcal{L}_{p_1 + p_2 + p_3 + p_4}
\)}
\vspace{-0.1cm}
\label{eq:add_loss}
\end{equation}
where $\mathcal{L}_{p_1}$, $\mathcal{L}_{p_2}$, $\mathcal{L}_{p_3}$, and $\mathcal{L}_{p_4}$ are the losses of each individual prediction maps. $\alpha = \beta = \gamma = \zeta = \delta = 1.0$ are the weights assigned to each loss.

%Following MERIT \cite{rahman2023multi}, we use the combinatorial loss aggregation strategy, MUTATION in all our experiments. Therefore, we compute the loss for $2^n-1$ combinatorial predictions synthesized from $n$ heads separately and then do a summation of them. We optimize this additive combinatorial loss during training.

\textbf{Output segmentation maps aggregation:} We consider the prediction map, $p_4$, from the last stage of our decoder as the final segmentation map. %summing up the weighted contributions of each segmentation map as in Equation \ref{eq:output_aggregation}:\vspace{-0.2cm}
%\begin{equation}
%   seg\_map = \alpha p_1 + \beta p_2 + \gamma p_3 + \zeta p_4
%    \vspace{-0.1cm}
%\label{eq:output_aggregation}
%\end{equation}
%where $\alpha = 0.5$, $\beta = 0.75$, $\gamma = 1.25$, and $\zeta = 2.0$ represent the weights assigned to each prediction head. 
Then, we obtain the final segmentation output by employing a $Sigmoid$ function for binary or a $Softmax$ function for multi-class segmentation.
%For all experiments, these weights are uniformly set to $1.0$. 
%\vspace{-.2cm}

%% file: sec/4_experiments.tex
\section{Experiments}
\label{sec:experiments}

\begin{figure}[t]
\begin{center}
%\fbox{\rule{0pt}{2in} \rule{.9\linewidth}{0pt}}
\includegraphics[width=0.85\linewidth]{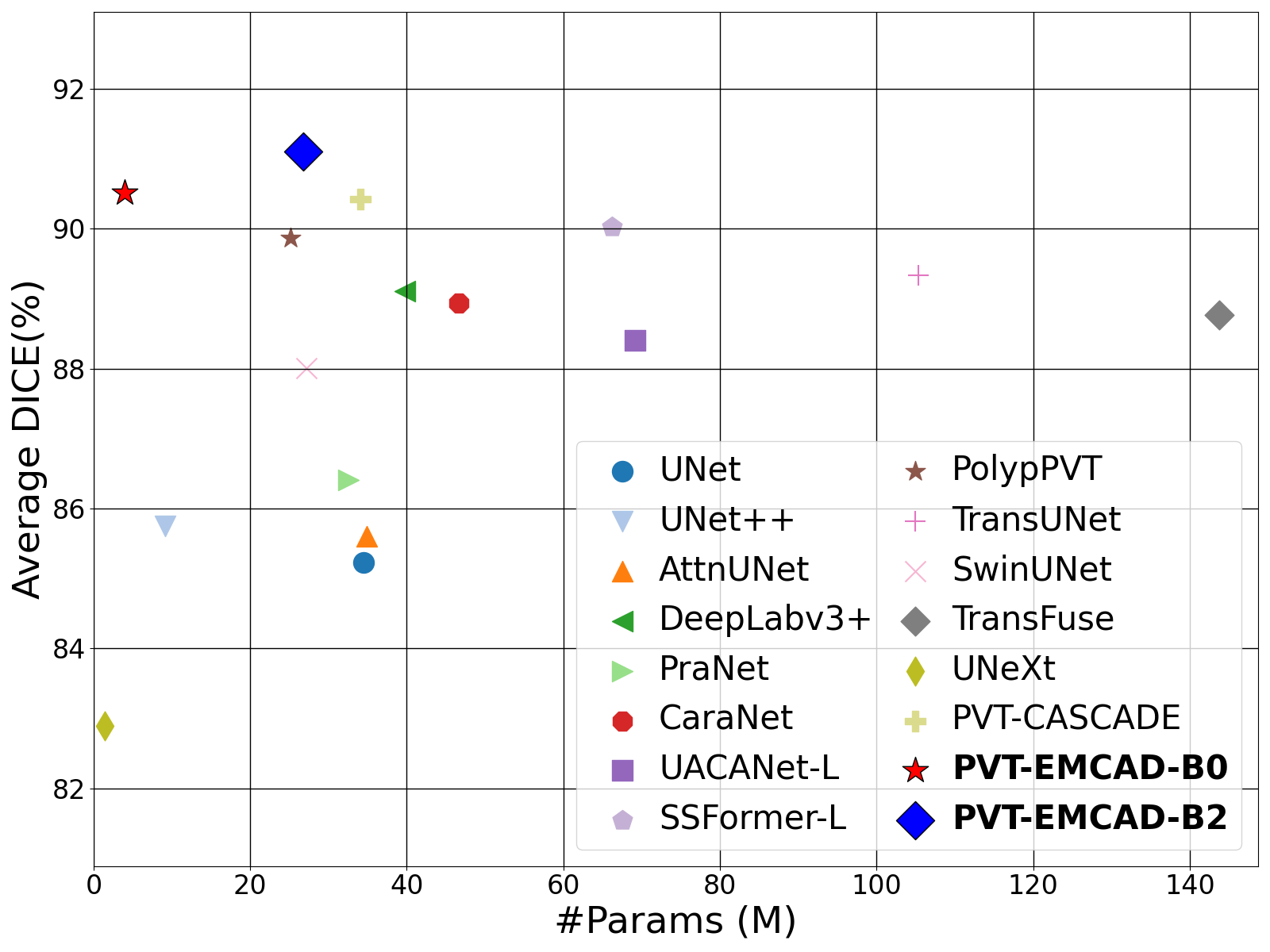}
\end{center}
\vspace{-0.7cm}
   \caption{Average DICE scores vs. \#Params for different methods over 10 binary medical image segmentation datasets. As shown, our proposed approaches (PVT-EMCAD-B0 and PVT-EMCAD-B2) have the fewest parameters, yet the highest DICE scores.}
   \vspace{-0.2cm}
\label{fig:dice_params}
\end{figure}

\begin{table*}[ht]
\begin{center}
\begin{adjustbox}{width=\textwidth}
\begin{tabular}{l|r|r|rrrrr|rr|rr|r|r}
\toprule
\multirow{2}{*}{Methods} & \multirow{2}{*}{\#Params} & \multirow{2}{*}{\#FLOPs} & \multicolumn{5}{c|}{Polyp} & \multicolumn{2}{c|}{Skin Lesion} & \multicolumn{2}{c|}{Cell} & \multirow{2}{*}{BUSI} & \multirow{2}{*}{Avg.}\\
\cline{4-12}
& & & Clinic & Colon & ETIS & Kvasir & BKAI & ISIC17 & ISIC18 & DSB18 & EM & & \\
\midrule
UNet \cite{ronneberger2015u} & 34.53M & 65.53G & 92.11 & 83.95 & 76.85 & 82.87 & 85.05 & 83.07 & 86.67 & 92.23 & 95.46 & 74.04 & 85.23 \\
UNet++ \cite{zhou2018unet++} & 9.16M & 34.65G & 92.17 & 87.88 & 77.40 & 83.36 & 84.07 & 82.98 & 87.46 & 91.97 & 95.48 & 74.76 & 85.75 \\
%UNet 3+ \cite{huang2020unet} & & & - & - & - & - & - & 84.31 & 86.83 & 90.83 & - & 74.69 \\
AttnUNet \cite{oktay2018attention} & 34.88M & 66.64G & 92.20 & 86.46 & 76.84 & 83.49 & 84.07 & 83.66 & 87.05 & 92.22 & \textbf{95.55} & 74.48 & 85.60 \\
%CENet \cite{gu2019net} & & & 91.53 & 83.11 & 75.03 & 84.92 & 81.62 & - & - & - & - & - \\
DeepLabv3+ \cite{chen2017deeplab} & 39.76M & 14.92G & 93.24 & 91.92 & 90.73 & 89.06 & 89.74 & 83.84 & 88.64 & 92.14 & 94.96 & 76.81 & 89.11 \\
%SegNet \cite{badrinarayanan2017segnet} & & & - & - & - & - & - & 85.58 & 89.86 & 91.06 & - & 76.84 \\
%SegFormer \cite{xie2021segformer} & & & 93.45 & 91.33 & 89.64 & 91.47 & 88.92 & 85.87 & 90.67 & 91.12 & - & 77.79 \\
PraNet \cite{fan2020pranet} & 32.55M & 6.93G & 91.71 & 89.16 & 83.84 & 84.82 & 85.56 & 83.03 & 88.56 & 89.89 & 92.37 & 75.14 & 86.41 \\
CaraNet \cite{lou2022caranet} & 46.64M & 11.48G & 94.08 & 91.19 & 90.25 & 89.74 & 89.71 & 85.02 & 90.18 & 89.15 & 92.78 & 77.34 & 88.94 \\
UACANet-L \cite{kim2021uacanet} & 69.16M & 31.51G & 94.16 & 91.02 & 89.77 & 90.17 & 90.35 & 83.72 & 89.76 & 88.86 & 89.28 & 76.96 & 88.41 \\
SSFormer-L \cite{wang2022stepwise} & 66.22M & 17.28G & 94.18 & 92.11 & 90.16 & 91.47 & 91.14 & 85.28 & 90.25 & 92.03 & 94.95 & 78.76 & 90.03  \\
PolypPVT \cite{dong2021polyp} & 25.11M & 5.30G & 94.13 & 91.53 & 89.93 & 91.56 & 91.17 & 85.56 & 90.36 & 90.69 & 94.40 & 79.35 & 89.87 \\
TransUNet \cite{chen2021transunet} & 105.32M & 38.52G & 93.90 & 91.63 & 87.79 & 91.08 & 89.17 & 85.00 & 89.16 & 92.04 & 95.27 & 78.30 & 89.33 \\
SwinUNet \cite{cao2021swin} & 27.17M  & 6.2G & 92.42 & 89.27 & 85.10 & 89.59 & 87.61 & 83.97 & 89.26 & 91.03 & 94.47 & 77.38 & 88.01 \\
TransFuse \cite{zhang2021transfuse} & 143.74M & 82.71G & 93.62 & 90.35 & 86.91 & 90.24 & 87.47 & 84.89 & 89.62 & 90.85 & 94.35 & 79.36 & 88.77  \\
UNeXt \cite{valanarasu2022unext} & 1.47M & 0.57G & 90.20 & 83.84 & 74.03 & 77.88 & 77.93 & 82.74 & 87.78 & 86.01 & 93.81 & 74.71 & 82.89 \\
PVT-CASCADE \cite{Rahman_2023_WACV} & 34.12M & 7.62G & 94.53 & 91.60 & 91.03 & 92.05 & 92.14 & 85.50 & 90.41 & 92.35 & 95.42 & 79.21 & 90.42 \\
%ACC-UNet \cite{ibtehaz2023acc} & 16.8 & 38 & - & - & - & - & - & 84.11 & 89.21 & 91.05 & 94.67 & 76.78 \\
%AAU-Net \cite{chen2022aau} & & & - & - & - & - & - & 83.97 & 88.87 & 91.10 & 94.54 & 76.98 \\
\midrule
PVT-EMCAD-B0 (\textbf{Ours}) & 3.92M & 0.84G & 94.60 & 91.71 & 91.65 & 91.95 & 91.30 & 85.67 & 90.70 & 92.46 & 95.35 & 79.80 & 90.52 \\
PVT-EMCAD-B2 (\textbf{Ours}) & 26.76M & 5.6G & \textbf{95.21} & \textbf{92.31} & \textbf{92.29} & \textbf{92.75} & \textbf{92.96} & \textbf{85.95} & \textbf{90.96} & \textbf{92.74} & 95.53 & \textbf{80.25} & \textbf{91.10} \\
\bottomrule
\end{tabular}
\end{adjustbox}
\vspace{-0.3cm}
\caption{Results of binary medical image segmentation (i.e., polyp, skin lesion, cell, and breast cancer). We reproduce the results of SOTA methods using their publicly available implementation with our train-val-test splits of 80:10:10. \#FLOPs of all the methods are reported for $256\times256$ inputs, except Swin-UNet ($224\times224$). All results are averaged over five runs. Best results are shown in bold.}
\vspace{-0.5cm}
\label{tab:results_binary_mis}
\end{center}
\end{table*}

\begin{table*}[]
\begin{center}
    {\small{
\begin{tabular}{l|rrr|rrrrrrrrr}
\toprule
\multirow{2}{*}{Architectures} & \multicolumn{3}{c|}{Average}                                                                                     & \multicolumn{1}{l}{\multirow{2}{*}{Aorta}} & \multicolumn{1}{l}{\multirow{2}{*}{GB}} & \multicolumn{1}{l}{\multirow{2}{*}{KL}} & \multicolumn{1}{l}{\multirow{2}{*}{KR}} & \multicolumn{1}{l}{\multirow{2}{*}{Liver}} & \multicolumn{1}{l}{\multirow{2}{*}{PC}} & \multicolumn{1}{l}{\multirow{2}{*}{SP}} & \multicolumn{1}{l}{\multirow{2}{*}{SM}} \\
                               & \multicolumn{1}{l}{DICE$\uparrow$} & \multicolumn{1}{l}{HD95$\downarrow$} & \multicolumn{1}{l|}{mIoU$\uparrow$} & \multicolumn{1}{l}{}                       & \multicolumn{1}{l}{}                    & \multicolumn{1}{l}{}                         & \multicolumn{1}{l}{}                         & \multicolumn{1}{l}{}                       & \multicolumn{1}{l}{}                    & \multicolumn{1}{l}{}                    & \multicolumn{1}{l}{}                    \\
\midrule
UNet \cite{ronneberger2015u}                   & 70.11                    & 44.69                    & 59.39                                        & 84.00                                      & 56.70                                   & 72.41                                        & 62.64                                        & 86.98                                      & 48.73                                   & 81.48                                   & 67.96                                   
\\
AttnUNet \cite{oktay2018attention}                   & 71.70                    & 34.47                    & 61.38                                   & 82.61                                      & 61.94                                   & 76.07                                        & 70.42                                        & 87.54                                      & 46.70                                   & 80.67                                   & 67.66                                   
\\
R50+UNet \cite{chen2021transunet}                   & 74.68                    & 36.87                    & $-$                                     & 84.18                                      & 62.84                                   & 79.19                                        & 71.29                                        & 93.35                                      & 48.23                                   & 84.41                                   & 73.92                                   
\\
R50+AttnUNet \cite{chen2021transunet}                   & 75.57                    & 36.97                    & $-$                                    & 55.92                                      & 63.91                                   & 79.20                                        & 72.71                                        & 93.56                                      & 49.37                                   & 87.19                                   & 74.95                                   
\\
SSFormer \cite{wang2022stepwise}                   & 78.01                    & 25.72                    & 67.23                                     & 82.78                                      & 63.74                                   & 80.72                                        & 78.11                                        & 93.53                                      & 61.53                                   & 87.07                                   & 76.61                                   \\
PolypPVT \cite{dong2021polyp}                       & 78.08                    & 25.61                    & 67.43                           & 82.34                                      & 66.14                                   & 81.21                                        & 73.78                                        & 94.37                                      & 59.34                                   & 88.05                                   & 79.4                                    \\
%TFCNs \cite{li2022tfcns}                         & 75.63                    & 30.63                    & 64.69                             & \textbf{88.23}                                      & 59.18                                   & 80.99                                        & 73.12                                        & 92.02                                      & 54.24                                   & 88.36                                   & 68.9                                    \\
TransUNet \cite{chen2021transunet}                     & 77.61                    & 26.9                     & 67.32                                  & 86.56                                      & 60.43                                   & 80.54                                        & 78.53                                        & 94.33                                      & 58.47                                   & 87.06                                   & 75.00                                      \\
SwinUNet \cite{cao2021swin}                      & 77.58                    & 27.32                    & 66.88                              & 81.76                                      & 65.95                                   & 82.32                                        & 79.22                                        & 93.73                                      & 53.81                                   & 88.04                                   & 75.79                                   \\
MT-UNet \cite{wang2022mixed}                      & 78.59                    & 26.59             & $-$       & 87.92                                      & 64.99                                   & 81.47                                        & 77.29                                        & 93.06                                     & 59.46                                  & 87.75                                   & 76.81                                  \\
MISSFormer \cite{huang2021missformer}                      & 81.96                    & 18.20             & $-$       & 86.99                                      &  68.65                                   &  85.21                                        & 82.00                                        & 94.41                                      & 65.67                                   & 91.92                                  & 80.81                                   \\ 
PVT-CASCADE \cite{Rahman_2023_WACV}                  & 81.06                    & 20.23                    & 70.88                                     & 83.01                                      & \textbf{70.59}                                   & 82.23                                        & 80.37                                        & 94.08                                      & 64.43                                   & 90.1                                    & 83.69                                   \\
TransCASCADE \cite{Rahman_2023_WACV}                & 82.68                    & 17.34                    & 73.48                                         & 86.63                                      & 68.48                                   & 87.66                                        & \textbf{84.56}                                        & 94.43                                     & 65.33                                   & 90.79                                   & 83.52                                   \\
%Cascaded MERIT \cite{rahman2023multi}                   & 84.32                    & 14.27    & 75.44                & 86.67                                      & 72.63                                   & 87.71                                       & 84.62                                        & 95.02                                      & \textbf{70.74}                                   &  \textbf{91.98}                                    & \textbf{85.17}                                   \\
%$\pm$0.2
%PVT-GCASCADE (\textbf{Ours})                 & 83.28                   &   15.83                  &  73.91                         &   86.50                                    &  71.71                                  &   87.07                                      &   83.77                                    &  95.31                                   &   66.72                                &   90.84                                & 83.58 \\
%MERIT-GCASCADE (\textbf{Ours})                 & \textbf{84.54}                   & \textbf{10.38}                    & \textbf{75.83}                           &  \textbf{88.05}                                     &    \textbf{74.81}                                & \textbf{88.01}                                        & \textbf{84.83}                                        & \textbf{95.38}                                      & 69.73                                   & 91.92                                   & 83.63 \\
\midrule
PVT-EMCAD-B0 (\textbf{Ours})                 & 81.97                   &   17.39                  &  72.64                         &   87.21                                    &  66.62                                  &   87.48                                      &   83.96                                    &  94.57                                   &   62.00                                &   \textbf{92.66}                                & 81.22
\\
PVT-EMCAD-B2 (\textbf{Ours})                 & \textbf{83.63}                   & \textbf{15.68}                    & \textbf{74.65}                           &  \textbf{88.14}                                     &    68.87                                & \textbf{88.08}                                        & 84.10                                        & \textbf{95.26}                                      & \textbf{68.51}                                   & 92.17                                   & \textbf{83.92}
\\
\bottomrule
\end{tabular}
}}
\end{center}
\vspace{-0.6cm}
\caption{Results of abdomen organ segmentation on Synapse Multi-organ dataset. DICE scores are reported for individual organs. Results of UNet, AttnUNet, PolypPVT, SSFormerPVT, TransUNet, and SwinUNet are taken from \cite{Rahman_2023_WACV}. $\uparrow$ ($\downarrow$) denotes the higher (lower) the better. `$-$' means missing data from the source. EMCAD results are averaged over five runs. Best results are shown in bold.}
\vspace{-0.4cm}
\label{tab:multi_organ_results}
\end{table*}

In this section, we present the details of our implementation followed by a comparative analysis of our PVT-EMCAD-B0 and PVT-EMCAD-B2 against SOTA methods. Datasets and evaluation metrics are in Supplementary Section \ref{asec:more_experiments}.

%We begin this section with details on the implementation. Subsequently, we present a comparative analysis of architectures based on our EMCAD decoder against SOTA methods to demonstrate the effectiveness of our proposed technique. Datasets and evaluation metrics are in supplementary \cref{sec:more_experiments}.

%an overview of the datasets used and the metrics for evaluation, followed by 

%In this section, we first describe the datasets and evaluation metrics, followed by implementation details. Then, we conduct a comparative analysis between our proposed G-CASCADE decoder-based architectures and SOTA methods to highlight the superior performance of our approach.

\begin{table}[]
\begin{center}
\begin{adjustbox}{width=0.48\textwidth}
\begin{tabular}{l|r|rrr}
\toprule
Methods        & Avg. DICE    & \multicolumn{1}{l}{RV} & \multicolumn{1}{l}{Myo} & \multicolumn{1}{l}{LV} \\
\midrule
R50+UNet   \cite{chen2021transunet}       & 87.55                        & 87.10                  & 80.63                   & 94.92                  \\
R50+AttnUNet  \cite{chen2021transunet}  & 86.75                        & 87.58                  & 79.20                   & 93.47                  \\
ViT+CUP \cite{chen2021transunet}   & 81.45                        & 81.46                  & 70.71                   & 92.18                 \\
R50+ViT+CUP \cite{chen2021transunet} & 87.57                        & 86.07                  & 81.88                   & 94.75                  \\
TransUNet  \cite{chen2021transunet}       & 89.71                        & 86.67                  & 87.27                   & 95.18                  \\
SwinUNet \cite{cao2021swin}         & 88.07                        & 85.77                  & 84.42                   & 94.03                  \\
%MT$-$UNet \cite{wang2022mixed}         & 90.43                        & 86.64                  & 89.04                   & 95.62                  \\
MT-UNet \cite{wang2022mixed}         & 90.43                        & 86.64                  & 89.04                   & 95.62                  \\
MISSFormer \cite{huang2021missformer}         & 90.86                        & 89.55                  & 88.04                   & 94.99                  \\
PVT-CASCADE \cite{Rahman_2023_WACV}      & 91.46                        & 89.97                   & 88.9                   & 95.50                   \\
TransCASCADE \cite{Rahman_2023_WACV}    & 91.63                       & 90.25                   &  89.14                  & 95.50 \\
Cascaded MERIT \cite{rahman2023multi}    & 91.85                       & 90.23                   &  89.53                  & 95.80 \\
%PVT-GCASCADE    & 91.95                        & 90.31                  & 89.63                   & 95.91 \\
%MERIT-GCASCADE    & \textbf{92.23}                        & \textbf{90.64}                  & \textbf{89.96}                   & \textbf{96.08} \\
\midrule
PVT-EMCAD-B0 (\textbf{Ours})    & 91.34$\pm0.2$                        & 89.37                  & 88.99                   & 95.65 \\
PVT-EMCAD-B2 (\textbf{Ours})    & \textbf{92.12$\pm0.2$}                        & \textbf{90.65}                  & \textbf{89.68}                   & \textbf{96.02} \\
\bottomrule %\\
\end{tabular}
\end{adjustbox}
\end{center}
\vspace{-0.5cm}
\caption{Results of cardiac organ segmentation on ACDC dataset. DICE scores (\%) are reported for individual organs. We get the results of SwinUNet from \cite{Rahman_2023_WACV}. Best results are shown in bold.} \vspace{-0.4cm}
\label{tab:acdc_results}
\end{table}

\subsection{Implementation details}
\label{ssec:impl_details}
We implement our network and conduct experiments using Pytorch 1.11.0 on a single NVIDIA RTX A6000 GPU with 48GB of memory. We utilize ImageNet \cite{deng2009imagenet} pre-trained PVTv2-b0 and PVTv2-b2 \cite{wang2022pvt} as encoders. In the MSDC of our decoder, we set the multi-scale kernels $[1,3,5]$ through an ablation study. We use the parallel arrangement of depth-wise convolutions in all experiments. Our models are trained using the AdamW optimizer \cite{loshchilov2017decoupled} with a learning rate and weight decay of $1e-4$. We generally train for 200 epochs with a batch size of 16, except for Synapse multi-organ (300 epochs, batch size 6) and ACDC cardiac organ (400 epochs, batch size 12), saving the best model based on the DICE score. We resize images to $352\times352$ and use a multi-scale \{0.75, 1.0, 1.25\} training strategy with a gradient clip limit of 0.5 for ClinicDB \cite{bernal2015wm}, Kvasir \cite{jha2020kvasir}, ColonDB \cite{vazquez2017benchmark}, ETIS \cite{vazquez2017benchmark}, BKAI \cite{ngoc2021neounet}, ISIC17 \cite{codella2018skin}, and ISIC18 \cite{codella2018skin}, while we resize images to $256\times256$ for BUSI \cite{al2020dataset}, EM \cite{cardona2010integrated}, and DSB18 \cite{caicedo2019nucleus}. For Synapse and ACDC datasets, images are resized to $224\times224$, with random rotation and flipping augmentations, optimizing a combined Cross-entropy (0.3) and DICE (0.7) loss. For binary segmentation, we utilize the combined weighted BinaryCrossEntropy (BCE) and weighted IoU loss function. 

%Following \cite{rahman2023cascade}, we resize images to $224\times224$, employ random rotation and flipping as data augmentation, and optimize the combined Cross-entropy (0.3) and DICE (0.7) loss for Synapse multi-organ and ACDC datasets. 
%We use a batch size of 6 and train each model for maximum of 300 epochs. We use the input resolution of $224 \times 224$ for PVT-GCASCADE and ($256\times256$, $224\times224$) for MERIT-GCASCADE. We apply random rotation and flipping for data augmentation. The combined weighted Cross-entropy (0.3) and DICE (0.7) loss are utilized as the loss function.

%\textbf{ACDC dataset.} For the ACDC dataset, we train each model for a maximum of 150 epochs with a batch size of 12. We set the input resolution as $224 \times 224$ for PVT-GCASCADE and ($256\times256$, $224\times224$) for MERIT-GCASCADE. We apply random flipping and rotation for data augmentation. We optimize the combined weighted Cross-entropy (0.3) and DICE (0.7) loss function.

%\textbf{Polyp datasets.} We resize the image to $352 \times 352$ and use a multi-scale \{0.75, 1.0, 1.25\} training strategy with a gradient clip limit of 0.5 like CASCADE \cite{Rahman_2023_WACV}. We use a batch size of 4 and train each model a maximum of 200 epochs. We optimize the combined weighted BinaryCrossEntropy (BCE) and weighted IoU loss function.

%\textbf{ISIC2018 dataset:} We resize the images into $384\times384$ resolution. Then, we train our model for 200 epochs with a batch size of 4 and a gradient clip of 0.5. We optimize the combined weighted BCE and weighted IoU loss function.

\vspace{-0.1cm}
\subsection{Results}
\label{ssec:results}
We compare our architectures (i.e., PVT-EMCAD-B0 and PVT-EMCAD-B2) with SOTA CNN and transformer-based segmentation methods on 12 datasets that belong to six medical image segmentation tasks. Qualitative results are in the Supplementary Section \ref{assec:qualitative_results}. %\cite{ronneberger2015u, oktay2018attention, chen2017deeplab, chen2021transunet, cao2021swin, huang2020unet, zhou2018unet++, badrinarayanan2017segnet, fan2020pranet, xie2021segformer, zhang2021transfuse, valanarasu2022unext, Rahman_2023_WACV, ibtehaz2023acc, chen2022aau} 
%on abdomen organ (i.e., Synapse multi-organ), cardiac organ (i.e., ACDC), skin lesion (i.e., ISIC17 \cite{codella2018skin}, ISIC18 \cite{codella2019skin}), polyp (i.e., ClinicDB \cite{bernal2015wm}, Kvasir \cite{jha2020kvasir}, ColonDB \cite{tajbakhsh2015automated}, ETIS \cite{}, BKAI \cite{}) datasets. The results of ISIC2018, polyp, and retinal vessel segmentation datasets are reported in the supplementary materials (Section B).

\vspace{-0.2cm}
\subsubsection{Results of binary medical image segmentation} 
Results for different methods on 10 binary medical image segmentation datasets are shown in Table \ref{tab:results_binary_mis} and Figure \ref{fig:dice_vs_flops}. Our PVT-EMCAD-B2 attains the highest average DICE score (91.10\%) with only 26.76M parameters and 5.6G FLOPs. The multi-scale depth-wise convolution in our EMCAD decoder, combined with the transformer encoder, contributes to these performance gains. %We believe that the use of efficient multi-scale depth-wise convolution in our EMCAD decoder together with the transformer encoder leads to these performance improvements. 

\textbf{Polyp segmentation:} Table \ref{tab:results_binary_mis} reveals that our PVT-EMCAD-B2 surpasses all SOTA methods in five polyp segmentation datasets. PVT-EMCAD-B2 achieves DICE score improvements of 1.08\%, 0.78\%, 2.36\%, 1.19\%, and 1.79\% over PolypPVT in ClinicDB, ColonDB, ETIS, Kvasir, and BKAI-IGI, despite having slightly more parameters and FLOPs. The smallest model UNeXt, exhibits the worst performance in all five polyp segmentation datasets. Our smaller model with only 3.92M parameters and 0.84G FLOPs also outperforms all the methods except PVT-CASCADE (in Kvasir and BKAI-IGH) and SSFormer-L (in ColonDB), which achieve the best performance among SOTA methods. In conclusion, our PVT-EMCAD-B2 achieves the new SOTA results in these five polyp segmentation datasets.

\textbf{Skin lesion segmentation:} Table \ref{tab:results_binary_mis} shows PVT-EMCAD-B2's strong performance on ISIC17 and ISIC18 skin lesion segmentation datasets, achieving DICE scores of 85.95\% and 90.96\%, surpassing DeepLabV3+ by 2.11\% and 2.32\%. It also beats the nearest method PVT-CASCADE by 0.45\% and 0.55\% in ISIC17 and ISIC18, respectively, though our decoder is significantly more efficient than CASCADE. Our PVT-EMCAD-B0 also shows huge potential in point care applications like skin lesion segmentation with only 3.92M parameters and 0.84G FLOPs.  

\textbf{Cell segmentation:} To evaluate our method's effectiveness in biological imaging, we use DSB18 \cite{caicedo2019nucleus} for cell nuclei and EM \cite{cardona2010integrated} for cell structure segmentation. As Table \ref{tab:results_binary_mis} indicates, our PVT-EMCAD-B2 sets a SOTA benchmark in cell nuclei segmentation on DSB18, outperforming DeepLabv3+, TransFuse, and PVT-CASCADE. On the EM dataset, PVT-EMCAD-B2 secures the second-best DICE score (95.53\%), offering significantly lower computational costs than the top-performing AttnUNet (95.55\%).
%We use the DSB18 for cell nuclei and EM for cell structure segmentation to assess the effectiveness of our method in biological imaging. Experimental results in Table \ref{tab:results_binary_mis} show that our PVT-EMCAD-B2 achieves SOTA performance by beating DeepLabv3+, TransFuse, and PVT-CASCADE by 1.71\%, 1.41\% and 0.22\%, respectively, in cell nuclei segmentation on the DSB18 dataset. In the EM dataset, although PVT-CASCADE shows the best result, our PVT-EMCAD-B2 achieves the second-best DICE score with significantly lower computational costs. 

\textbf{Breast cancer segmentation:} We conduct experiments on the BUSI dataset for breast cancer segmentation in ultrasound images. Our PVT-EMCAD-B2 achieves the SOTA DICE score (80.25\%) on this dataset. Furthermore, our PVT-EMCAD-B0 outperforms the computationally similar method UNeXt by a notable margin of 5.54\%. %We have conducted experiments on the BUSI dataset to segment breast cancer in breast ultrasound images. Our PVT-EMCAD-B2 achieves the SOTA DICE score (80.25\%) on this dataset. Our PVT-EMCAD-B0 outperforms the computationally similar method UNeXt by 5.03\%.  
\vspace{-0.2cm}

\begin{table}[]
\begin{center}
    %{\small{
    \begin{adjustbox}{width=0.48\textwidth}
\begin{tabular}{cccrrrr}
\toprule
\multicolumn{3}{c}{Components} & \multicolumn{2}{c}{\#FLOPs(G)} & \multicolumn{1}{c}{\#Params} & \multicolumn{1}{r}{Avg}  \\
Cascaded     & LGAG     & MSCAM   & \multicolumn{1}{c}{224} & \multicolumn{1}{c}{256}& \multicolumn{1}{c}{(M)} & DICE                    \\
\midrule
No           & No     & No     &   0 & 0 & 0 & 80.10$\pm$0.2     \\
Yes          & No     & No     &  0.100 & 0.131 & 0.224 & 81.08$\pm$0.2    \\
Yes          & Yes    & No     & 0.108 & 0.141 & 0.235 & 81.92$\pm$0.2          \\
Yes          & No     & Yes    &  0.373 & 0.487 & 1.898 & 82.86$\pm$0.3      \\
Yes          & Yes    & Yes    & 0.381 & 0.498 & 1.91 & \textbf{83.63$\pm$0.3}       \\
\bottomrule 
\end{tabular}
%}}
\end{adjustbox}
\end{center}
\vspace{-0.5cm}
\caption{Effect of different components of EMCAD with PVTv2-b2 encoder on Synapse multi-organ dataset. \#FLOPs are reported for input resolution of $224\times224$ and $256\times256$. All results are averaged over five runs. Best results are shown in bold.}
\vspace{-0.5cm}
\label{tab:ablation_components}
\end{table}

\begin{table*}[t]
\centering 
    {%\small{
\begin{adjustbox}{width=0.92\textwidth}
{\begin{tabular}{lrrrrrrrrrr}
\toprule
Conv. kernels & $[1]$ & $[3]$ & $[5]$ & $[1,3]$ & $[3,3]$ & $[1,3,5]$ & $[3,3,3]$ & $[3,5,7]$ & $[1,3,5,7]$ & $[1,3,5,7,9]$\\
\midrule
Synapse & 82.43 & 82.79 & 82.74 & 82.98 & 82.81 & \textbf{83.63} & 82.92 & 83.11 & 83.57 & 83.34 \\
ClinicDB &  94.81 &  94.90 &  94.98 & 95.13 &  95.06 & \textbf{95.21} &  95.15 & 95.03 & 95.18 & 95.07 \\
\bottomrule %\\
\end{tabular}}
\end{adjustbox}

}%}
\vspace{-0.2cm}
\caption{Effect of multi-scale kernels in the depth-wise convolution of MSDC on ClinicDB and Synapse multi-organ datasets. We use the PVTv2-b2 encoder for these experiments. All results are averaged over five runs. Best results are highlighted in bold.}
\vspace{-0.4cm}
\label{tab:multi_kernels}
\end{table*}
\subsubsection{Results of abdomen organ segmentation}
Table \ref{tab:multi_organ_results} shows that our PVT-EMCAD-B2 excels in abdomen organ segmentation on the Synapse multi-organ dataset, achieving the highest average DICE score of 83.63\% and surpassing all SOTA CNN- and transformer-based methods. It outperforms PVT-CASCADE by 2.57\% in DICE score and 4.55 in HD95 distance, indicating superior organ boundary location. Our EMCAD decoder boosts individual organ segmentation, significantly outperforming SOTA methods on six of eight organs.
%We can also see a boost in the DICE scores of individual organ segmentation by our EMCAD decoder. 
\vspace{-0.4cm}
\subsubsection{Results of cardiac organ segmentation}
Table \ref{tab:acdc_results}
shows the DICE scores of our PVT-EMCAD-B2 and PVT-EMCAD-B0 along with other SOTA methods, on the MRI images of the ACDC dataset for cardiac organ segmentation. Our PVT-EMCAD-B2 achieves the highest average DICE score of 92.12\%, thus improving about 0.27\% over Cascaded MERIT though our network has significantly lower computational cost. Besides, PVT-EMCAD-B2 has better DICE scores in all three organ segmentations.

%% file: sec/5_ablation_study.tex
\vspace{-0.1cm}
\section{Ablation Studies}
\label{sec:ablation_study}

\begin{table}[t]
\centering 
    {%\small{
\begin{adjustbox}{width=0.48\textwidth}
{\begin{tabular}{lcrrr}
\toprule
Encoders & Decoders    &\multicolumn{1}{c}{\#FLOPs(G)} & \multicolumn{1}{c}{\#Params(M)} & \multicolumn{1}{l}{DICE (\%)} \\
\midrule
   PVTv2-B0      & CASCADE &  0.439 &  2.32 &  80.54 \\
   PVTv2-B0    & EMCAD (\textbf{Ours})    & \textbf{0.110} &  \textbf{0.507} & \textbf{81.97}   \\
\midrule 
  PVTv2-B2     & CASCADE &  1.93 & 9.27 & 82.78     \\
  PVTv2-B2   & EMCAD (\textbf{Ours}) & \textbf{0.381} & \textbf{1.91}     & \textbf{83.63}   \\
\bottomrule %\\
\end{tabular}}
\end{adjustbox}

}%}
\vspace{-0.2cm}
\caption{Comparison with the baseline decoder on Synapse Multi-organ dataset. We only report the \#FLOPs (with input resolution of $224\times224$) and the \#parameters of the decoders. All the results are averaged over five runs. Best results are shown in bold.}
\vspace{-0.5cm}
\label{tab:compare_baseline_decoder}
\end{table}
%------------------------------------------------------------------------- 

%\subsection{Transfer learning experiments}
%\subsection{Different encoders}
%\subsection{Small model and decoder}
%\subsection{Compare with other decoders}

In this section, we conduct ablation studies to explore different aspects of our architectures and the experimental framework. More ablations are in Supplementary Section \ref{asec:ablation}.

% from other paper
%In Table 4, the CASCADE performs the best with the MiT. We believe that this is because the convolution operation inside MiT can maintain the consistency of the spatial information of the model. Furthermore, the experiments in Table 5 demonstrate the effectiveness of the CASCADE and its components.

%\textbf{Effective enhancement/refinement of features.}
%We visualize the features of our CASCADE, as well as Cascaded Upsampler (CUP) \cite{chen2021transunet} in Figure \ref{fig:middle_features}. We compute the average of all channels in the feature map and then produce the heatmap using OpenCV-Python. It is evident from Figure \ref{fig:middle_features} that the attention mechanism used in our CASCADE helps identify, enhance, and group the pixels better than CUP.  

\vspace{-0.1cm}
\subsection{Effect of different components of EMCAD}
We conduct a set of experiments on the Synapse multi-organ dataset to understand the effect of different components of our EMCAD decoder. We start with only the encoder and add different modules such as Cascaded structure, LGAG, and MSCAM to understand their effect. Table \ref{tab:ablation_components} exhibits that the cascaded structure of the decoder helps to improve performance over the non-cascaded one. The incorporation of LGAG and MSCAM improves performance, however, MSCAM proves to be more effective. When both the LGAG and MSCAM modules are used together, it produces the best DICE score of 83.63\%. It is also evident that there is about 3.53\% improvement in the DICE score with an additional 0.381G FLOPs and 1.91M parameters.   

\vspace{-0.1cm}
\subsection{Effect of multi-scale kernels in MSCAM}
We have conducted another set of experiments on Synapse multi-organ and ClinicDB datasets to understand the effect of different multi-scale kernels used for depth-wise convolutions in MSDC. Table \ref{tab:multi_kernels} reports these results which show that performance improves from $1\times1$ to $3\times3$ kernel. When $1\times1$ kernel is used together with $3\times3$ it improves more than when using them alone. However, when two $3\times3$ kernels are used together, performance drops. The incorporation of a $5\times5$ kernel with $1\times1$ and $3\times3$ kernels further improves the performance and it achieves the best results in both Synapse multi-organ and ClinicDB datasets. If we add additional larger kernels (e.g., $7\times7$, $9\times9$), the performance of both datasets drops. Based on these empirical observations, we choose $[1,3,5]$ kernels in all our experiments. %Based on these empirical observations, we choose $[1,3,5]$ kernels for all the binary segmentation tasks and $[1,3,5,7]$ kernels for multi-class segmentation tasks. %Nonetheless, it improves the performance of the Synapse multi-organ dataset. If we have another large kernel, it reduces the performances of both datasets.   
 
\vspace{-0.1cm}
\subsection{Comparison with the baseline decoder}

In Table \ref{tab:compare_baseline_decoder}, we report the experimental results with the computational complexity of our EMCAD decoder and a baseline decoder, namely CASCADE. From Table \ref{tab:compare_baseline_decoder}, we can see that our EMCAD decoder with PVTv2-b2 requires 80.3\% fewer FLOPs and 79.4\% fewer parameters to outperform (by 0.85\%) the respective CASCADE decoder. Similarly, our EMCAD decoder with PVTv2-B0 achieves 1.43\% better DICE score than the CASCADE decoder with 78.1\% fewer parameters and 74.9\% fewer FLOPs.        

%\subsection{Qualitative results}

%% file: sec/6_conclusion.tex
\vspace{-0.2cm}
\section{Conclusions}
\label{sec:conclusion}

In this paper, we have presented EMCAD, a new and efficient multi-scale convolutional attention decoder designed for multi-stage feature aggregation and refinement in medical image segmentation. EMCAD employs a multi-scale depth-wise convolution block, which is key for capturing diverse scale information within feature maps, a critical factor for precision in medical image segmentation. This design choice, using depth-wise convolutions instead of standard $3\times3$ convolution blocks, makes EMCAD notably efficient. 

Our experiments reveal that EMCAD surpasses the recent CASCADE decoder in DICE scores with 79.4\% fewer parameters and 80.3\% less FLOPs. Our extensive experiments also confirm EMCAD's superior performance compared to SOTA methods across 12 public datasets covering six different 2D medical image segmentation tasks. EMCAD's compatibility with smaller encoders makes it an excellent fit for point-of-care applications while maintaining high performance. We anticipate that our EMCAD decoder will be a valuable asset in enhancing a variety of medical image segmentation and semantic segmentation tasks.

%It also employs a new grouped attention gate for feature aggregation with the skip connections by using $3\times3$ group convolutions instead of $1\times1$ point-wise convolution. 

%In this paper, we have introduced a new efficient multi-scale convolutional attention decoder, namely EMCAD, for multi-stage feature aggregation and refinement. EMCAD utilizes a multi-scale depth-wise convolution block that enables capturing multi-scale information in the feature maps which is crucial for accurate medical image segmentation. Due to using depth-wise convolution blocks instead of basic $3\times3$ convolution blocks, EMCAD is computationally \textit{very efficient}. Our experimental results show that EMACD outperforms a recent decoder, CASCADE, in DICE scores with 79.4\% fewer parameters and 80.3\% fewer FLOPs. When EMCAD is used with tiny encoders, it can easily be used with point-of-care applications with high performance. Our experimental results also demonstrate the superiority of our EMCAD decoder over SOTA methods on twelve public medical image segmentation datasets that belong to six 2D medical image segmentation tasks. Finally, we believe that our proposed decoder will improve other downstream medical image segmentation and semantic segmentation tasks.

%\vspace{-0.2cm}
\noindent \textbf{Acknowledgements:}
This work is supported in part by the NSF grant CNS 2007284, and in part by the iMAGiNE Consortium (https://imagine.utexas.edu/).
\vspace{-0.2cm}

%% file: sec/X_suppl.tex
\clearpage
\setcounter{page}{1}
\maketitlesupplementary

\begin{figure*}[]%[H]%[t]
  \centering
  %\vspace{-0.3cm}
  %\fbox{\rule{0pt}{0.5in} \rule{0.9\linewidth}{0pt}}
  \includegraphics[width=0.95\linewidth]{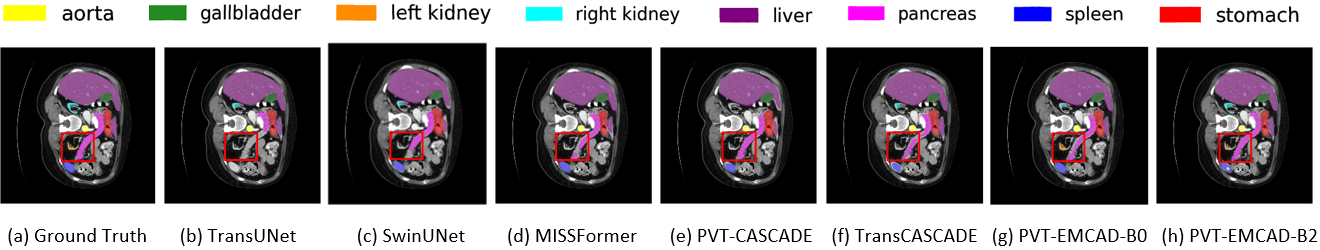}
  %\vspace{-0.65cm}
   \caption{Qualitative results of multi-organ segmentation on Synapse Multi-organ dataset. The red rectangular box highlights incorrectly segmented organs by SOTA methods.}
   \label{fig:qualitative_synapse}
   %\vspace{-0.45cm}
\end{figure*}

\begin{figure*}%[H]%[t]
  \centering
  %\vspace{-0.3cm}
  %\fbox{\rule{0pt}{0.5in} \rule{0.9\linewidth}{0pt}}
  \includegraphics[width=0.75\linewidth]{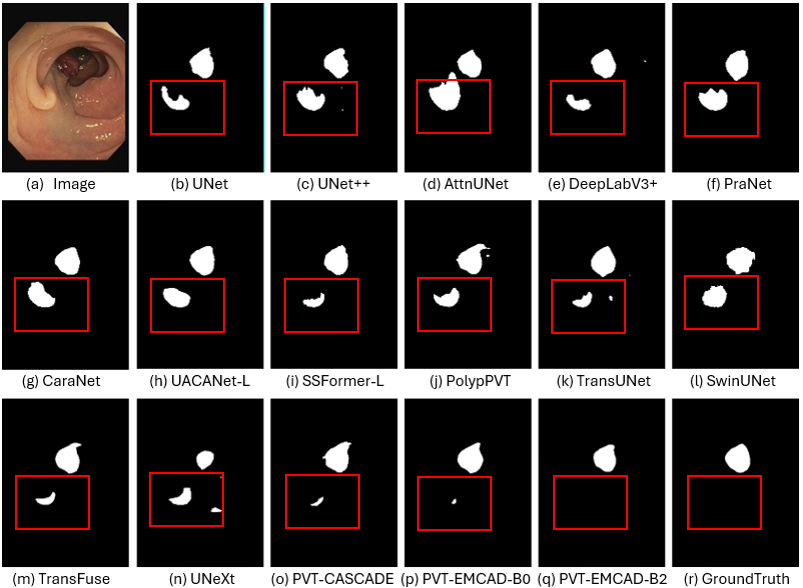}
  %\vspace{-0.65cm}
   \caption{Qualitative results of polyp segmentation. The red rectangular box highlights incorrectly segmented polyps by SOTA methods.}% and multi-organ segmentation (right).}
   \label{fig:qualitative_polyp}
   %\vspace{-0.45cm}
\end{figure*}

\section{Experimental Details}
\label{asec:more_experiments}
This section extends our Section \ref{sec:experiments} in the original paper by describing the datasets and evaluation metrics, followed by additional experimental results.

\subsection{Datasets}
\label{assec:dataset_metrics}
%\vspace{-.1cm}
To evaluate the performance of our EMCAD decoder, we carry out experiments across 12 datasets that belong to six medical image segmentation tasks, as described next. %We present the description of abdomen organ, cardiac organ, polyp, Skin lesion, cell, and breast cancer segmentation datasets below.%, and retinal vessel segmentation datasets are described in supplementary materials (Section A).

\noindent \textbf{Polyp segmentation:} We use five polyp segmentation datasets: Kvasir \cite{jha2020kvasir} (1,000 images), ClinicDB \cite{bernal2015wm} (612 images), ColonDB \cite{vazquez2017benchmark}
(379 images), ETIS \cite{vazquez2017benchmark} (196 images), and BKAI \cite{ngoc2021neounet} (1,000 images). These datasets contain images from different imaging centers/clinics, having greater diversity in image nature as well as size and shape of polyps. 

\noindent \textbf{Abdomen organ segmentation:} We use the Synapse multi-organ dataset\footnote{\href{https://www.synapse.org/\#!Synapse:syn3193805/wiki/217789}{https://www.synapse.org/\#!Synapse:syn3193805/wiki/217789 }} for abdomen organ segmentation. This dataset contains 30 abdominal CT scans which have 3,779 axial contrast-enhanced slices. Each CT scan has 85-198 slices of $512 \times 512$ pixels.
Following TransUNet \cite{chen2021transunet}, we use the same 18 scans for training (2,212 axial slices) and 12 scans for validation. We segment only eight abdominal organs, namely aorta, gallbladder (GB), left kidney (KL), right kidney (KR), liver, pancreas (PC), spleen (SP), and stomach (SM). 

\noindent \textbf{Cardiac organ segmentation:} We use ACDC dataset\footnote{\href{https://www.creatis.insa-lyon.fr/Challenge/acdc/}{https://www.creatis.insa-lyon.fr/Challenge/acdc/}} for cardiac organ segmentation. It contains 100 cardiac MRI scans having three sub-organs, namely right ventricle (RV), myocardium (Myo), and left ventricle (LV). Following TransUNet \cite{chen2021transunet}, we use 70 cases (1,930 axial slices) for training, 10 for validation, and 20 for testing. 

\noindent \textbf{Skin lesion segmentation:} We use ISIC17 \cite{codella2018skin} (2,000 training, 150 validation, and 600 testing images) and ISIC18 \cite{codella2019skin} (2,594 images) for skin lesion segmentation. 

\noindent \textbf{Breast cancer segmentation:} We use BUSI \cite{al2020dataset} dataset for breast cancer segmentation. Following \cite{valanarasu2022unext}, we use 647 (437 benign and 210 malignant) images from this dataset. 

\noindent \textbf{Cell nuclei/structure segmentation:} We use the DSB18 \cite{caicedo2019nucleus} (670 images) and EM \cite{cardona2010integrated} (30 images) datasets of biological imaging for cell nuclei/structure segmentation. 

We use a train-val-test split of 80:10:10 in ClinicDB, Kvasir, ColonDB, ETIS, BKAI, ISIC18, DSB18, EM, and BUSI datasets. For ISIC17, we use the official train-val-test sets provided by the competition organizer.  
% and MT\_UNet \cite{wang2022mixed}

\subsection{Evaluation metrics} 
\label{assec:eval_metrics}
We use the DICE score to evaluate performance on all the datasets. However, we also use 95\% Hausdorff Distance (HD95) and mIoU as additional evaluation metrics for Synapse multi-organ segmentation. The DICE score $DSC(Y, P)$, $IoU(Y, P)$, and HD95 distance $D_H(Y, P)$ are calculated using Equations \ref{eq:dice}, \ref{eq:iou}, and \ref{eq:95hd}, respectively:
%(95th percentile of the distances between boundary points in $Y$ and $\hat{Y}$) 
\begin{equation}\label{eq:dice}
DSC(Y, P) = \frac{2 \times \lvert Y \cap P \rvert}{\lvert Y \rvert + \lvert P \rvert}\times100
\end{equation}
\begin{equation}\label{eq:iou}
IoU(Y, P) = \frac{\lvert Y \cap P \rvert}{\lvert Y \cup P \rvert}\times100
\end{equation}
%\vspace{-0.4cm}
\begin{equation}\label{eq:95hd}
\small
D_H(Y, P) = \max \{\max_{y \in Y} \min_{p \in P}d(y, p), \{\max_{p \in P} \min_{y \in Y}d(y, p)\}
%\max\{d_{Y\hat{Y}},d_{\hat{Y}Y}\} = 
\end{equation}
where $Y$ and $P$ are the ground truth and predicted segmentation map, respectively. 

%Therefore, following the experimental scheme of Polyp-PVT \cite{dong2021polyp}, we train and test PVT-CASCADE on the Kavsir-SEG [8] and CVC-ClinicDB [1] benchmark datasets, respectively, to assess the accurate prediction and learning ability of models in the Kavsir-SEG and CVC-ClinicDB test set, respectively. 
\vspace{0.4cm}
\subsection{Qualitative results} 
\label{assec:qualitative_results}
This subsection describes the qualitative results of different methods including our EMCAD. From, the qualitative results on Synapse Multi-organ dataset in  Figure \ref{fig:qualitative_synapse}, we can see that most of the methods face challenges segmenting the left kidney (orange) and part of the pancreas (pink). However, our PVT-EMCAD-B0 (Figure \ref{fig:qualitative_synapse}g) and PVT-EMCAD-B2 (Figure \ref{fig:qualitative_synapse}h) can segment those organs more accurately (see red rectangular box) with significantly lower computational costs. Similarly, qualitative results of polyp segmentation on a representative image from ClinicDB dataset in Figure \ref{fig:qualitative_polyp} show that predicted segmentation outputs of our PVT-EMCAD-B0 (Figure \ref{fig:qualitative_polyp}p) and PVT-EMCAD-B2 (Figure \ref{fig:qualitative_polyp}q) have strong overlaps with the GroundTruth mask (Figure \ref{fig:qualitative_polyp}r), while existing SOTA methods exhibit false segmentation of polyp (see red rectangular box).  

%\subsubsection{Results of generalizability experiments}
%\label{asec:generalizability_results}

\vspace{0.4cm}
\section{Additional Ablation Study}
\label{asec:ablation}
%This section extends \cref{sec:ablation_study} by describing three more ablation studies related to our architectural design and experimental setup. 
This section further elaborates on Section \ref{sec:ablation_study} by detailing five additional ablation studies related to our architectural design and experimental setup.
%\subsection{Effectiveness of GAG over AG}

%\subsection{Attention scores visualization of different components}
\begin{table}[]
\centering 
    {%\small{
\begin{adjustbox}{width=0.48\textwidth}
{\begin{tabular}{lcrrrr}
\toprule
Architectures & Depth-wise convolutions & \multicolumn{1}{r}{Synapse} & \multicolumn{1}{r}{ClinicDB}\\
\midrule
PVT-EMCAD-B0 & Sequential & 81.82$\pm$0.3 & 94.57$\pm$0.2
\\ PVT-EMCAD-B0 & Parallel  & \textbf{81.97$\pm$0.2} & \textbf{94.60$\pm$0.2} \\
\midrule
PVT-EMCAD-B2 & Sequential  & 83.54$\pm$0.3  & 95.15$\pm$0.3
\\ PVT-EMCAD-B2 &  Parallel  & \textbf{83.63$\pm$0.2}  & \textbf{95.21$\pm$0.2} \\
\bottomrule %\\
\end{tabular}}
\end{adjustbox}

}%}
\vspace{-0.1cm}
\caption{Results of parallel and sequential depth-wise convolution in MSDC on Synapse multi-organ and ClinicDB datasets. All results are averaged over five runs. Best results are in bold.}
\vspace{-0.3cm}
\label{tab:arrangements_parallel_seq}
\end{table}

\begin{table}[]
\centering 
    {%\small{
\begin{adjustbox}{width=0.48\textwidth}
{\begin{tabular}{lcrrr}
\toprule
Architectures & Module  & \multicolumn{1}{r}{Params(K)} & \multicolumn{1}{r}{FLOPs(M)} &\multicolumn{1}{r}{Synapse}\\
\midrule
PVT-EMCAD-B0 & AG & 31.62 & 15.91 & 81.74
\\ PVT-EMCAD-B0 & \textbf{LGAG}  & \textbf{5.51} & \textbf{5.24} & \textbf{81.97}\\
\midrule
PVT-EMCAD-B2 & AG  & 124.68  & 61.68 & 83.51
\\ PVT-EMCAD-B2 &  \textbf{LGAG}  & \textbf{11.01}  & \textbf{10.47} & \textbf{83.63} \\
\bottomrule %\\
\end{tabular}}
\end{adjustbox}

}%}
\vspace{-0.2cm}
\caption{LGAG vs. AG (Attention gate) \cite{oktay2018attention} on Synapse multi-organ dataset. The total \#Params and \#FLOPs of three AG/LGAGs in our decoder are reported for an input resolution of $256\times256$. All results are averaged over five runs. Best results are in bold.}
\vspace{-0.3cm}
\label{tab:lgag_vs_ag}
\end{table}

\subsection{Parallel vs. sequential depth-wise convolution}
\label{assec:parallel_seq}
We have conducted another set of experiments to decide whether we use multi-scale depth-wise convolutions in parallel or sequential. Table \ref{tab:arrangements_parallel_seq} presents the results of these experiments which show that there is no significant impact of the arrangements though the parallel convolutions provide a slightly improved performance (0.03\% to 0.15\%). We also observe higher standard deviations among runs in the case of sequential convolutions. Hence, in all our experiments, we use multi-scale depth-wise convolutions \textit{in parallel}. 

\begin{table*}[]
\begin{center}
    {%\small{
\begin{adjustbox}{width=0.98\textwidth}
{\begin{tabular}{lc|rrr|rrrrrrrrr}
\toprule
\multirow{2}{*}{Architectures} & \multirow{2}{*}{Pretrain} & \multicolumn{3}{c|}{Average}                                                                                     & \multicolumn{1}{l}{\multirow{2}{*}{Aorta}} & \multicolumn{1}{l}{\multirow{2}{*}{GB}} & \multicolumn{1}{l}{\multirow{2}{*}{KL}} & \multicolumn{1}{l}{\multirow{2}{*}{KR}} & \multicolumn{1}{l}{\multirow{2}{*}{Liver}} & \multicolumn{1}{l}{\multirow{2}{*}{PC}} & \multicolumn{1}{l}{\multirow{2}{*}{SP}} & \multicolumn{1}{l}{\multirow{2}{*}{SM}} \\
                              & & \multicolumn{1}{l}{DICE$\uparrow$} & \multicolumn{1}{l}{HD95$\downarrow$} & \multicolumn{1}{l|}{mIoU$\uparrow$} & \multicolumn{1}{l}{} & \multicolumn{1}{l}{} & \multicolumn{1}{l}{} & \multicolumn{1}{l}{} & \multicolumn{1}{l}{} & \multicolumn{1}{l}{} & \multicolumn{1}{l}{} & \multicolumn{1}{l}{}      \\
\midrule %+0.2
PVT-EMCAD-B0     & No     & 77.47 &   19.93  &  66.72 &   81.96 &  \textbf{69.41}  &   83.88  &   74.82 &  93.45  &   54.41 &   88.97 & 72.85 \\
PVT-EMCAD-B0      & Yes     & \textbf{81.97} &   \textbf{17.39}  &  \textbf{72.64}   &   \textbf{87.21}   &  66.62  & \textbf{87.48}  &   \textbf{83.96} &  \textbf{94.57} &   \textbf{62.00}   &   \textbf{92.66}   & \textbf{81.22} \\
\midrule
PVT-EMCAD-B2         & No        & 80.18   & 18.83   & 70.21  &  85.98   &    68.10  & 84.62   & 79.93  & 93.96   & 61.61  & 90.99   & 76.23
\\
PVT-EMCAD-B2      & Yes   & \textbf{83.63}  & \textbf{15.68}   & \textbf{74.65}  &  \textbf{88.14}  &    \textbf{68.87}  & \textbf{88.08} & \textbf{84.10} & \textbf{95.26} & \textbf{68.51}  & \textbf{92.17} & \textbf{83.92} \\
\bottomrule
\end{tabular}}
\end{adjustbox}
}%}
\end{center}
%\vspace{-0.4cm}
\caption{Effect of transfer learning from ImageNet pre-trained weights on Synapse multi-organ dataset. $\uparrow$ ($\downarrow$) denotes the higher (lower) the better. All results are averaged over five runs. Best results are in bold.}
%\vspace{-0.1cm}
\label{tab:transfer_learing}
\end{table*}

\begin{table}%[H]
\centering
%\vspace{-0.2cm}
\begin{adjustbox}{width=1\linewidth}

\begin{tabular}{l|r|r|r|r|r|r|r}
\hline
        DS     & EM    & BUSI  & Clinic  & Kvasir  & ISIC18 & Synapse & ACDC  \\ \hline
No           & 95.74 & 79.64 & 94.96  & 92.51  & 90.74  & 82.03   & 92.08 \\ 
Yes          & 95.53 & 80.25 & 95.21  & 92.75  & 90.96  & 83.63   & 92.12 \\ \hline
\end{tabular}
\end{adjustbox}
%\vspace{-0.2cm}
\caption{Effect of deep supervision (DS). PVT-EMCAD-B2 with DS achieves slightly better DICE scores in 6 out of 7 datasets.}
%\vspace{-0.4cm}
\label{tab:ds_ablation}
\end{table}

\begin{table}[t]
\centering 
    {%\small{
\begin{adjustbox}{width=0.48\textwidth}
{\begin{tabular}{lrrrrr}
\toprule
Architectures & Resolutions  & \multicolumn{1}{r}{FLOPs(G)} & \multicolumn{1}{r}{DICE}\\ %&
%\midrule
%Parallel MERIT \cite{rahman2023multi} & $256\times256$ & 33.31 &  84.22
%\\ Cascaded MERIT \cite{rahman2023multi} & $256\times256$  & 33.31 &  84.90 \\

\midrule
PVT-EMCAD-B0  & $224\times224$ & 0.64 & 81.97
\\ PVT-EMCAD-B0 & $256\times256$  & 0.84 &  82.63 \\
PVT-EMCAD-B0  & $384\times384$  &  1.89 & 84.81 \\
PVT-EMCAD-B0  & $512\times512$  & 3.36 & 85.52 \\
\midrule
PVT-EMCAD-B2 & $224\times224$  & 4.29  & 83.63
\\
PVT-EMCAD-B2  &  $256\times256$ & 5.60  & 84.47 \\
PVT-EMCAD-B2  &  $384\times384$ &  12.59  & 85.78 \\
PVT-EMCAD-B2  &  $512\times512$ &  22.39 & 86.53 \\
\bottomrule %\\
\end{tabular}}
\end{adjustbox}

}%}
\vspace{-0.2cm}
\caption{Effect of input resolutions on Synapse multi-organ dataset. All results are averaged over five runs.}
\vspace{-0.2cm}
\label{tab:effect_input_resolutions}
\end{table}

\subsection{Effectiveness of our large-kernel grouped attention gate (LGAG) over attention gate (AG)}
\label{assec:lgag_vs_ag}
Table \ref{tab:lgag_vs_ag} presents experimental results of EMCAD with original AG \cite{oktay2018attention} and our LGAG. We can conclude that our LGAG achieves better DICE scores with significant reductions in \#Params (82.57\% for PVT-EMCAD-B0 and 91.17\% for PVT-EMCAD-B2) and \#FLOPs (67.06\% for PVT-EMCAD-B0 and 83.03\% for PVT-EMCAD-B2) than AG. The reduction in \#Params and \#FLOPs is bigger for the larger models. Therefore, our LGAG demonstrates improved scalability with models that have a greater number of channels, yielding enhanced DICE scores.   
%Therefore, our LGAG scales better with the models that have a larger number of channels, yielding enhanced DICE scores. 

\subsection{Effect of transfer learning from ImageNet pre-trained weights}
\label{asec:transfer_learing}
We conduct experiments on the Synapse multi-organ dataset to show the effect of transfer learning from the ImageNet pre-trained encoder. Table \ref{tab:transfer_learing} reports the results of these experiments which show that transfer learning from ImageNet pre-trained PVT-v2 encoders significantly boosts the performance. Specifically, for PVT-EMCAD-B0, the DICE, mIoU, and HD95 scores are improved by 4.5\%, 5.92\%, and 2.54, respectively. Likewise, for PVT-EMCAD-B2, the DICE, mIoU, and HD95 scores are improved by 3.45\%, 4.44\%, and 3.15, respectively. We can also conclude that transfer learning has a comparatively greater impact on the smaller PVT-EMCAD-B0 model than the larger PVT-EMCAD-B2 model. For individual organs, transfer learning significantly boosts the performance of all organ segmentation, except the Gallbladder (GB).  

\subsection{Effect of deep supervision}
\label{asec:effect_deep_supervision}
We have conducted an ablation study that drops the Deep Supervision (DS). Results of our PVT-EMCAD-B2 on seven datasets are given in Table \ref{tab:ds_ablation}. Our PVT-EMCAD-B2 with DS achieves slightly better DICE scores in six out of seven datasets. Among all the datasets, the DS has the largest impact on the Synapse Multi-organ dataset.    

\subsection{Effect of input resolutions}
\label{asec:effect_input_resolutions}
Table \ref{tab:effect_input_resolutions} presents the results of our PVT-EMCAD-B0 and PVT-EMCAD-B2 architectures with different input resolutions. From this table, it is evident that the DICE scores improve with the increase in input resolution. However, these improvements in DICE score come with the increment in \#FLOPs. Our PVT-EMCAD-B0 achieves an 85.52\% DICE score with only 3.36G FLOPs when using $512\times512$ inputs. On the other hand, our PVT-EMCAD-B2 achieves the best DICE score (86.53\%) with 22.39G FLOPs when using $512\times512$ inputs. We also observe that our PVT-EMCAD-B2 with 5.60G FLOPs when using $256\times256$ inputs shows a 1.05\% lower DICE score than PVT-EMCAD-B0 with 3.36G FLOPs. Therefore, we can conclude that PVT-EMCAD-B0 is more suitable for larger input resolutions than PVT-EMCAD-B2.

%% file: main.bbl
\begin{thebibliography}{62}
\providecommand{\natexlab}[1]{#1}
\providecommand{\url}[1]{\texttt{#1}}
\expandafter\ifx\csname urlstyle\endcsname\relax
  \providecommand{\doi}[1]{doi: #1}\else
  \providecommand{\doi}{doi: \begingroup \urlstyle{rm}\Url}\fi

\bibitem[Al-Dhabyani et~al.(2020)Al-Dhabyani, Gomaa, Khaled, and Fahmy]{al2020dataset}
Walid Al-Dhabyani, Mohammed Gomaa, Hussien Khaled, and Aly Fahmy.
\newblock Dataset of breast ultrasound images.
\newblock \emph{Data in brief}, 28:\penalty0 104863, 2020.

\bibitem[Badrinarayanan et~al.(2017)Badrinarayanan, Kendall, and Cipolla]{badrinarayanan2017segnet}
Vijay Badrinarayanan, Alex Kendall, and Roberto Cipolla.
\newblock Segnet: A deep convolutional encoder-decoder architecture for image segmentation.
\newblock \emph{IEEE Trans. Pattern Anal. Mach. Intell.}, 39\penalty0 (12):\penalty0 2481--2495, 2017.

\bibitem[Bernal et~al.(2015)Bernal, S{\'a}nchez, Fern{\'a}ndez-Esparrach, Gil, Rodr{\'\i}guez, and Vilari{\~n}o]{bernal2015wm}
Jorge Bernal, F~Javier S{\'a}nchez, Gloria Fern{\'a}ndez-Esparrach, Debora Gil, Cristina Rodr{\'\i}guez, and Fernando Vilari{\~n}o.
\newblock Wm-dova maps for accurate polyp highlighting in colonoscopy: Validation vs. saliency maps from physicians.
\newblock \emph{Comput. Med. Imaging Graph.}, 43:\penalty0 99--111, 2015.

\bibitem[Caicedo et~al.(2019)Caicedo, Goodman, Karhohs, Cimini, Ackerman, Haghighi, Heng, Becker, Doan, McQuin, et~al.]{caicedo2019nucleus}
Juan~C Caicedo, Allen Goodman, Kyle~W Karhohs, Beth~A Cimini, Jeanelle Ackerman, Marzieh Haghighi, CherKeng Heng, Tim Becker, Minh Doan, Claire McQuin, et~al.
\newblock Nucleus segmentation across imaging experiments: the 2018 data science bowl.
\newblock \emph{Nature methods}, 16\penalty0 (12):\penalty0 1247--1253, 2019.

\bibitem[Cao et~al.(2021)Cao, Wang, Chen, Jiang, Zhang, Tian, and Wang]{cao2021swin}
Hu Cao, Yueyue Wang, Joy Chen, Dongsheng Jiang, Xiaopeng Zhang, Qi Tian, and Manning Wang.
\newblock Swin-unet: Unet-like pure transformer for medical image segmentation.
\newblock \emph{arXiv preprint arXiv:2105.05537}, 2021.

\bibitem[Cardona et~al.(2010)Cardona, Saalfeld, Preibisch, Schmid, Cheng, Pulokas, Tomancak, and Hartenstein]{cardona2010integrated}
Albert Cardona, Stephan Saalfeld, Stephan Preibisch, Benjamin Schmid, Anchi Cheng, Jim Pulokas, Pavel Tomancak, and Volker Hartenstein.
\newblock An integrated micro-and macroarchitectural analysis of the drosophila brain by computer-assisted serial section electron microscopy.
\newblock \emph{PLoS biology}, 8\penalty0 (10):\penalty0 e1000502, 2010.

\bibitem[Chen et~al.(2022)Chen, Li, Dai, Zhang, and Yap]{chen2022aau}
Gongping Chen, Lei Li, Yu Dai, Jianxun Zhang, and Moi~Hoon Yap.
\newblock Aau-net: an adaptive attention u-net for breast lesions segmentation in ultrasound images.
\newblock \emph{IEEE Trans. Med. Imaging}, 2022.

\bibitem[Chen et~al.(2021)Chen, Lu, Yu, Luo, Adeli, Wang, Lu, Yuille, and Zhou]{chen2021transunet}
Jieneng Chen, Yongyi Lu, Qihang Yu, Xiangde Luo, Ehsan Adeli, Yan Wang, Le Lu, Alan~L Yuille, and Yuyin Zhou.
\newblock Transunet: Transformers make strong encoders for medical image segmentation.
\newblock \emph{arXiv preprint arXiv:2102.04306}, 2021.

\bibitem[Chen et~al.(2017{\natexlab{a}})Chen, Zhang, Xiao, Nie, Shao, Liu, and Chua]{chen2017sca}
Long Chen, Hanwang Zhang, Jun Xiao, Liqiang Nie, Jian Shao, Wei Liu, and Tat-Seng Chua.
\newblock Sca-cnn: Spatial and channel-wise attention in convolutional networks for image captioning.
\newblock In \emph{IEEE Conf. Comput. Vis. Pattern Recog.}, pages 5659--5667, 2017{\natexlab{a}}.

\bibitem[Chen et~al.(2017{\natexlab{b}})Chen, Papandreou, Kokkinos, Murphy, and Yuille]{chen2017deeplab}
Liang-Chieh Chen, George Papandreou, Iasonas Kokkinos, Kevin Murphy, and Alan~L Yuille.
\newblock Deeplab: Semantic image segmentation with deep convolutional nets, atrous convolution, and fully connected crfs.
\newblock \emph{IEEE Trans. Pattern Anal. Mach. Intell.}, 40\penalty0 (4):\penalty0 834--848, 2017{\natexlab{b}}.

\bibitem[Chen et~al.(2018{\natexlab{a}})Chen, Zhu, Papandreou, Schroff, and Adam]{chen2018encoder}
Liang-Chieh Chen, Yukun Zhu, George Papandreou, Florian Schroff, and Hartwig Adam.
\newblock Encoder-decoder with atrous separable convolution for semantic image segmentation.
\newblock In \emph{Eur. Conf. Comput. Vis.}, pages 801--818, 2018{\natexlab{a}}.

\bibitem[Chen et~al.(2018{\natexlab{b}})Chen, Tan, Wang, and Hu]{chen2018reverse}
Shuhan Chen, Xiuli Tan, Ben Wang, and Xuelong Hu.
\newblock Reverse attention for salient object detection.
\newblock In \emph{Eur. Conf. Comput. Vis.}, pages 234--250, 2018{\natexlab{b}}.

\bibitem[Chu et~al.(2021)Chu, Tian, Zhang, Wang, Wei, Xia, and Shen]{chu2021conditional}
Xiangxiang Chu, Zhi Tian, Bo Zhang, Xinlong Wang, Xiaolin Wei, Huaxia Xia, and Chunhua Shen.
\newblock Conditional positional encodings for vision transformers.
\newblock \emph{arXiv preprint arXiv:2102.10882}, 2021.

\bibitem[Codella et~al.(2019)Codella, Rotemberg, Tschandl, Celebi, Dusza, Gutman, Helba, Kalloo, Liopyris, Marchetti, et~al.]{codella2019skin}
Noel Codella, Veronica Rotemberg, Philipp Tschandl, M~Emre Celebi, Stephen Dusza, David Gutman, Brian Helba, Aadi Kalloo, Konstantinos Liopyris, Michael Marchetti, et~al.
\newblock Skin lesion analysis toward melanoma detection 2018: A challenge hosted by the international skin imaging collaboration (isic).
\newblock \emph{arXiv preprint arXiv:1902.03368}, 2019.

\bibitem[Codella et~al.(2018)Codella, Gutman, Celebi, Helba, Marchetti, Dusza, Kalloo, Liopyris, Mishra, Kittler, et~al.]{codella2018skin}
Noel~CF Codella, David Gutman, M~Emre Celebi, Brian Helba, Michael~A Marchetti, Stephen~W Dusza, Aadi Kalloo, Konstantinos Liopyris, Nabin Mishra, Harald Kittler, et~al.
\newblock Skin lesion analysis toward melanoma detection: A challenge at the 2017 international symposium on biomedical imaging (isbi), hosted by the international skin imaging collaboration (isic).
\newblock In \emph{IEEE Int. Symp. Biomed. Imaging}, pages 168--172. IEEE, 2018.

\bibitem[Deng et~al.(2009)Deng, Dong, Socher, Li, Li, and Fei-Fei]{deng2009imagenet}
Jia Deng, Wei Dong, Richard Socher, Li-Jia Li, Kai Li, and Li Fei-Fei.
\newblock Imagenet: A large-scale hierarchical image database.
\newblock In \emph{IEEE Conf. Comput. Vis. Pattern Recog.}, pages 248--255. Ieee, 2009.

\bibitem[Dong et~al.(2021)Dong, Wang, Fan, Li, Fu, and Shao]{dong2021polyp}
Bo Dong, Wenhai Wang, Deng-Ping Fan, Jinpeng Li, Huazhu Fu, and Ling Shao.
\newblock Polyp-pvt: Polyp segmentation with pyramid vision transformers.
\newblock \emph{arXiv preprint arXiv:2108.06932}, 2021.

\bibitem[Dosovitskiy et~al.(2020)Dosovitskiy, Beyer, Kolesnikov, Weissenborn, Zhai, Unterthiner, Dehghani, Minderer, Heigold, Gelly, et~al.]{dosovitskiy2020image}
Alexey Dosovitskiy, Lucas Beyer, Alexander Kolesnikov, Dirk Weissenborn, Xiaohua Zhai, Thomas Unterthiner, Mostafa Dehghani, Matthias Minderer, Georg Heigold, Sylvain Gelly, et~al.
\newblock An image is worth 16x16 words: Transformers for image recognition at scale.
\newblock \emph{arXiv preprint arXiv:2010.11929}, 2020.

\bibitem[et~al.(2021)]{isensee2021nnu}
Isensee et al.
\newblock nnu-net: a self-configuring method for deep learning-based biomedical image segmentation.
\newblock \emph{Nature methods}, 18\penalty0 (2):\penalty0 203--211, 2021.

\bibitem[Fan et~al.(2020)Fan, Ji, Zhou, Chen, Fu, Shen, and Shao]{fan2020pranet}
Deng-Ping Fan, Ge-Peng Ji, Tao Zhou, Geng Chen, Huazhu Fu, Jianbing Shen, and Ling Shao.
\newblock Pranet: Parallel reverse attention network for polyp segmentation.
\newblock In \emph{Int. Conf. Med. Image Comput. Comput. Assist. Interv.}, pages 263--273. Springer, 2020.

\bibitem[He et~al.(2016)He, Zhang, Ren, and Sun]{he2016deep}
Kaiming He, Xiangyu Zhang, Shaoqing Ren, and Jian Sun.
\newblock Deep residual learning for image recognition.
\newblock In \emph{IEEE Conf. Comput. Vis. Pattern Recog.}, pages 770--778, 2016.

\bibitem[Howard et~al.(2017)Howard, Zhu, Chen, Kalenichenko, Wang, Weyand, Andreetto, and Adam]{howard2017mobilenets}
Andrew~G Howard, Menglong Zhu, Bo Chen, Dmitry Kalenichenko, Weijun Wang, Tobias Weyand, Marco Andreetto, and Hartwig Adam.
\newblock Mobilenets: Efficient convolutional neural networks for mobile vision applications.
\newblock \emph{arXiv preprint arXiv:1704.04861}, 2017.

\bibitem[Hu et~al.(2018)Hu, Shen, and Sun]{hu2018squeeze}
Jie Hu, Li Shen, and Gang Sun.
\newblock Squeeze-and-excitation networks.
\newblock In \emph{IEEE Conf. Comput. Vis. Pattern Recog.}, pages 7132--7141, 2018.

\bibitem[Huang et~al.(2020)Huang, Lin, Tong, Hu, Zhang, Iwamoto, Han, Chen, and Wu]{huang2020unet}
Huimin Huang, Lanfen Lin, Ruofeng Tong, Hongjie Hu, Qiaowei Zhang, Yutaro Iwamoto, Xianhua Han, Yen-Wei Chen, and Jian Wu.
\newblock Unet 3+: A full-scale connected unet for medical image segmentation.
\newblock In \emph{ICASSP}, pages 1055--1059. IEEE, 2020.

\bibitem[Huang et~al.(2021)Huang, Deng, Li, and Yuan]{huang2021missformer}
Xiaohong Huang, Zhifang Deng, Dandan Li, and Xueguang Yuan.
\newblock Missformer: An effective medical image segmentation transformer.
\newblock \emph{arXiv preprint arXiv:2109.07162}, 2021.

\bibitem[Ibtehaz and Kihara(2023)]{ibtehaz2023acc}
Nabil Ibtehaz and Daisuke Kihara.
\newblock Acc-unet: A completely convolutional unet model for the 2020s.
\newblock In \emph{Int. Conf. Med. Image Comput. Comput. Assist. Interv.}, pages 692--702. Springer, 2023.

\bibitem[Ioffe and Szegedy(2015)]{ioffe2015batch}
Sergey Ioffe and Christian Szegedy.
\newblock Batch normalization: Accelerating deep network training by reducing internal covariate shift.
\newblock In \emph{Int. Conf. Mach. Learn.}, pages 448--456. pmlr, 2015.

\bibitem[Islam et~al.(2020)Islam, Jia, and Bruce]{islam2020much}
Md~Amirul Islam, Sen Jia, and Neil~DB Bruce.
\newblock How much position information do convolutional neural networks encode?
\newblock \emph{arXiv preprint arXiv:2001.08248}, 2020.

\bibitem[Jha et~al.(2020)Jha, Smedsrud, Riegler, Halvorsen, Lange, Johansen, and Johansen]{jha2020kvasir}
Debesh Jha, Pia~H Smedsrud, Michael~A Riegler, P{\aa}l Halvorsen, Thomas~de Lange, Dag Johansen, and H{\aa}vard~D Johansen.
\newblock Kvasir-seg: A segmented polyp dataset.
\newblock In \emph{Int. Conf. Multimedia Model.}, pages 451--462. Springer, 2020.

\bibitem[Kim et~al.(2021)Kim, Lee, and Kim]{kim2021uacanet}
Taehun Kim, Hyemin Lee, and Daijin Kim.
\newblock Uacanet: Uncertainty augmented context attention for polyp segmentation.
\newblock In \emph{ACM Int. Conf. Multimedia}, pages 2167--2175, 2021.

\bibitem[Krizhevsky and Hinton(2010)]{krizhevsky2010convolutional}
Alex Krizhevsky and Geoff Hinton.
\newblock Convolutional deep belief networks on cifar-10.
\newblock \emph{Unpublished manuscript}, 40\penalty0 (7):\penalty0 1--9, 2010.

\bibitem[Krizhevsky et~al.(2012)Krizhevsky, Sutskever, and Hinton]{krizhevsky2012imagenet}
Alex Krizhevsky, Ilya Sutskever, and Geoffrey~E Hinton.
\newblock Imagenet classification with deep convolutional neural networks.
\newblock \emph{Adv. Neural Inform. Process. Syst.}, 25, 2012.

\bibitem[Lin et~al.(2023)Lin, Yan, Deng, Zheng, and Yu]{lin2023convformer}
Xian Lin, Zengqiang Yan, Xianbo Deng, Chuansheng Zheng, and Li Yu.
\newblock Convformer: Plug-and-play cnn-style transformers for improving medical image segmentation.
\newblock In \emph{Int. Conf. Med. Image Comput. Comput. Assist. Interv.}, pages 642--651. Springer, 2023.

\bibitem[Liu et~al.(2021)Liu, Lin, Cao, Hu, Wei, Zhang, Lin, and Guo]{liu2021swin}
Ze Liu, Yutong Lin, Yue Cao, Han Hu, Yixuan Wei, Zheng Zhang, Stephen Lin, and Baining Guo.
\newblock Swin transformer: Hierarchical vision transformer using shifted windows.
\newblock In \emph{Int. Conf. Comput. Vis.}, pages 10012--10022, 2021.

\bibitem[Liu et~al.(2022)Liu, Mao, Wu, Feichtenhofer, Darrell, and Xie]{liu2022convnet}
Zhuang Liu, Hanzi Mao, Chao-Yuan Wu, Christoph Feichtenhofer, Trevor Darrell, and Saining Xie.
\newblock A convnet for the 2020s.
\newblock In \emph{IEEE Conf. Comput. Vis. Pattern Recog.}, pages 11976--11986, 2022.

\bibitem[Loshchilov and Hutter(2017)]{loshchilov2017decoupled}
Ilya Loshchilov and Frank Hutter.
\newblock Decoupled weight decay regularization.
\newblock \emph{arXiv preprint arXiv:1711.05101}, 2017.

\bibitem[Lou et~al.(2021)Lou, Guan, and Loew]{lou2021dc}
Ange Lou, Shuyue Guan, and Murray Loew.
\newblock Dc-unet: rethinking the u-net architecture with dual channel efficient cnn for medical image segmentation.
\newblock In \emph{Med. Imaging 2021: Image Process.}, pages 758--768. SPIE, 2021.

\bibitem[Lou et~al.(2022)Lou, Guan, Ko, and Loew]{lou2022caranet}
Ange Lou, Shuyue Guan, Hanseok Ko, and Murray~H Loew.
\newblock Caranet: context axial reverse attention network for segmentation of small medical objects.
\newblock In \emph{Med. Imaging 2022: Image Process.}, pages 81--92. SPIE, 2022.

\bibitem[Nair and Hinton(2010)]{nair2010rectified}
Vinod Nair and Geoffrey~E Hinton.
\newblock Rectified linear units improve restricted boltzmann machines.
\newblock In \emph{Int. Conf. Mach. Learn.}, pages 807--814, 2010.

\bibitem[Ngoc~Lan et~al.(2021)Ngoc~Lan, An, Hang, Long, Trung, Thuy, and Sang]{ngoc2021neounet}
Phan Ngoc~Lan, Nguyen~Sy An, Dao~Viet Hang, Dao~Van Long, Tran~Quang Trung, Nguyen~Thi Thuy, and Dinh~Viet Sang.
\newblock Neounet: Towards accurate colon polyp segmentation and neoplasm detection.
\newblock In \emph{Adv. Vis. Comput. – Int. Symp.}, pages 15--28. Springer, 2021.

\bibitem[Oktay et~al.(2018)Oktay, Schlemper, Folgoc, Lee, Heinrich, Misawa, Mori, McDonagh, Hammerla, Kainz, et~al.]{oktay2018attention}
Ozan Oktay, Jo Schlemper, Loic~Le Folgoc, Matthew Lee, Mattias Heinrich, Kazunari Misawa, Kensaku Mori, Steven McDonagh, Nils~Y Hammerla, Bernhard Kainz, et~al.
\newblock Attention u-net: Learning where to look for the pancreas.
\newblock \emph{arXiv preprint arXiv:1804.03999}, 2018.

\bibitem[Rahman and Marculescu(2023{\natexlab{a}})]{Rahman_2023_WACV}
Md~Mostafijur Rahman and Radu Marculescu.
\newblock Medical image segmentation via cascaded attention decoding.
\newblock In \emph{IEEE/CVF Winter Conf. Appl. Comput. Vis.}, pages 6222--6231, 2023{\natexlab{a}}.

\bibitem[Rahman and Marculescu(2023{\natexlab{b}})]{rahman2023multi}
Md~Mostafijur Rahman and Radu Marculescu.
\newblock Multi-scale hierarchical vision transformer with cascaded attention decoding for medical image segmentation.
\newblock In \emph{Med. Imaging Deep Learn.}, 2023{\natexlab{b}}.

\bibitem[Ronneberger et~al.(2015)Ronneberger, Fischer, and Brox]{ronneberger2015u}
Olaf Ronneberger, Philipp Fischer, and Thomas Brox.
\newblock U-net: Convolutional networks for biomedical image segmentation.
\newblock In \emph{Int. Conf. Med. Image Comput. Comput. Assist. Interv.}, pages 234--241. Springer, 2015.

\bibitem[Sandler et~al.(2018)Sandler, Howard, Zhu, Zhmoginov, and Chen]{sandler2018mobilenetv2}
Mark Sandler, Andrew Howard, Menglong Zhu, Andrey Zhmoginov, and Liang-Chieh Chen.
\newblock Mobilenetv2: Inverted residuals and linear bottlenecks.
\newblock In \emph{IEEE Conf. Comput. Vis. Pattern Recog.}, pages 4510--4520, 2018.

\bibitem[Simonyan and Zisserman(2014)]{simonyan2014very}
Karen Simonyan and Andrew Zisserman.
\newblock Very deep convolutional networks for large-scale image recognition.
\newblock \emph{arXiv preprint arXiv:1409.1556}, 2014.

\bibitem[Szegedy et~al.(2015)Szegedy, Liu, Jia, Sermanet, Reed, Anguelov, Erhan, Vanhoucke, and Rabinovich]{szegedy2015going}
Christian Szegedy, Wei Liu, Yangqing Jia, Pierre Sermanet, Scott Reed, Dragomir Anguelov, Dumitru Erhan, Vincent Vanhoucke, and Andrew Rabinovich.
\newblock Going deeper with convolutions.
\newblock In \emph{IEEE Conf. Comput. Vis. Pattern Recog.}, pages 1--9, 2015.

\bibitem[Tan and Le(2019)]{tan2019efficientnet}
Mingxing Tan and Quoc Le.
\newblock Efficientnet: Rethinking model scaling for convolutional neural networks.
\newblock In \emph{Int. Conf. Mach. Learn.}, pages 6105--6114. PMLR, 2019.

\bibitem[Tu et~al.(2022)Tu, Talebi, Zhang, Yang, Milanfar, Bovik, and Li]{tu2022maxvit}
Zhengzhong Tu, Hossein Talebi, Han Zhang, Feng Yang, Peyman Milanfar, Alan Bovik, and Yinxiao Li.
\newblock Maxvit: Multi-axis vision transformer.
\newblock In \emph{Eur. Conf. Comput. Vis.}, pages 459--479. Springer, 2022.

\bibitem[Valanarasu and Patel(2022)]{valanarasu2022unext}
Jeya Maria~Jose Valanarasu and Vishal~M Patel.
\newblock Unext: Mlp-based rapid medical image segmentation network.
\newblock In \emph{Int. Conf. Med. Image Comput. Comput. Assist. Interv.}, pages 23--33. Springer, 2022.

\bibitem[V{\'a}zquez et~al.(2017)V{\'a}zquez, Bernal, S{\'a}nchez, Fern{\'a}ndez-Esparrach, L{\'o}pez, Romero, Drozdzal, and Courville]{vazquez2017benchmark}
David V{\'a}zquez, Jorge Bernal, F~Javier S{\'a}nchez, Gloria Fern{\'a}ndez-Esparrach, Antonio~M L{\'o}pez, Adriana Romero, Michal Drozdzal, and Aaron Courville.
\newblock A benchmark for endoluminal scene segmentation of colonoscopy images.
\newblock \emph{J. Healthc. Eng.}, 2017, 2017.

\bibitem[Wang et~al.(2022{\natexlab{a}})Wang, Cao, Wang, and Zaiane]{wang2022uctransnet}
Haonan Wang, Peng Cao, Jiaqi Wang, and Osmar~R Zaiane.
\newblock Uctransnet: rethinking the skip connections in u-net from a channel-wise perspective with transformer.
\newblock In \emph{AAAI}, pages 2441--2449, 2022{\natexlab{a}}.

\bibitem[Wang et~al.(2022{\natexlab{b}})Wang, Xie, Lin, Iwamoto, Han, Chen, and Tong]{wang2022mixed}
Hongyi Wang, Shiao Xie, Lanfen Lin, Yutaro Iwamoto, Xian-Hua Han, Yen-Wei Chen, and Ruofeng Tong.
\newblock Mixed transformer u-net for medical image segmentation.
\newblock In \emph{ICASSP}, pages 2390--2394. IEEE, 2022{\natexlab{b}}.

\bibitem[Wang et~al.(2022{\natexlab{c}})Wang, Huang, Tang, Meng, Su, and Song]{wang2022stepwise}
Jinfeng Wang, Qiming Huang, Feilong Tang, Jia Meng, Jionglong Su, and Sifan Song.
\newblock Stepwise feature fusion: Local guides global.
\newblock \emph{arXiv preprint arXiv:2203.03635}, 2022{\natexlab{c}}.

\bibitem[Wang et~al.(2021)Wang, Xie, Li, Fan, Song, Liang, Lu, Luo, and Shao]{wang2021pyramid}
Wenhai Wang, Enze Xie, Xiang Li, Deng-Ping Fan, Kaitao Song, Ding Liang, Tong Lu, Ping Luo, and Ling Shao.
\newblock Pyramid vision transformer: A versatile backbone for dense prediction without convolutions.
\newblock In \emph{Int. Conf. Comput. Vis.}, pages 568--578, 2021.

\bibitem[Wang et~al.(2022{\natexlab{d}})Wang, Xie, Li, Fan, Song, Liang, Lu, Luo, and Shao]{wang2022pvt}
Wenhai Wang, Enze Xie, Xiang Li, Deng-Ping Fan, Kaitao Song, Ding Liang, Tong Lu, Ping Luo, and Ling Shao.
\newblock Pvt v2: Improved baselines with pyramid vision transformer.
\newblock \emph{Comput. Vis. Media}, 8\penalty0 (3):\penalty0 415--424, 2022{\natexlab{d}}.

\bibitem[Woo et~al.(2018)Woo, Park, Lee, and Kweon]{woo2018cbam}
Sanghyun Woo, Jongchan Park, Joon-Young Lee, and In~So Kweon.
\newblock Cbam: Convolutional block attention module.
\newblock In \emph{Eur. Conf. Comput. Vis.}, pages 3--19, 2018.

\bibitem[Xie et~al.(2021)Xie, Wang, Yu, Anandkumar, Alvarez, and Luo]{xie2021segformer}
Enze Xie, Wenhai Wang, Zhiding Yu, Anima Anandkumar, Jose~M Alvarez, and Ping Luo.
\newblock Segformer: Simple and efficient design for semantic segmentation with transformers.
\newblock \emph{Adv. Neural Inform. Process. Syst.}, 34:\penalty0 12077--12090, 2021.

\bibitem[Yu et~al.(2022)Yu, Luo, Zhou, Si, Zhou, Wang, Feng, and Yan]{yu2022metaformer}
Weihao Yu, Mi Luo, Pan Zhou, Chenyang Si, Yichen Zhou, Xinchao Wang, Jiashi Feng, and Shuicheng Yan.
\newblock Metaformer is actually what you need for vision.
\newblock In \emph{IEEE Conf. Comput. Vis. Pattern Recog.}, pages 10819--10829, 2022.

\bibitem[Zhang et~al.(2018)Zhang, Zhou, Lin, and Sun]{zhang2018shufflenet}
Xiangyu Zhang, Xinyu Zhou, Mengxiao Lin, and Jian Sun.
\newblock Shufflenet: An extremely efficient convolutional neural network for mobile devices.
\newblock In \emph{IEEE Conf. Comput. Vis. Pattern Recog.}, pages 6848--6856, 2018.

\bibitem[Zhang et~al.(2021)Zhang, Liu, and Hu]{zhang2021transfuse}
Yundong Zhang, Huiye Liu, and Qiang Hu.
\newblock Transfuse: Fusing transformers and cnns for medical image segmentation.
\newblock In \emph{Int. Conf. Med. Image Comput. Comput. Assist. Interv.}, pages 14--24. Springer, 2021.

\bibitem[Zhou et~al.(2018)Zhou, Rahman~Siddiquee, Tajbakhsh, and Liang]{zhou2018unet++}
Zongwei Zhou, Md~Mahfuzur Rahman~Siddiquee, Nima Tajbakhsh, and Jianming Liang.
\newblock Unet++: A nested u-net architecture for medical image segmentation.
\newblock In \emph{Deep Learn. Med. Image Anal. Multimodal Learn. Clin. Decis. Support}, pages 3--11. Springer, 2018.

\end{thebibliography}
